\documentclass[12pt]{article} 

\usepackage{geometry}
\usepackage{verbatim}
\usepackage{latexsym}
\usepackage{graphics}
\usepackage{epsfig} 
\usepackage{amsfonts}
\usepackage{fancyhdr}

\renewenvironment{quotation}{%
   \list{}{%
     \leftmargin0.5cm   % this is the adjusting screw
     \rightmargin\leftmargin
   }
   \item\relax
}
{\endlist}

\def\bvert{{$\vert$}} 
\def\g{\,\vert \,} 
\def\ev{\,\mbox{ev}\,} 
\def\rr{\noindent - }  

\begin{document}

% \title{A Sharper Image: 	The Quest for Knowledge and Recursive Production of Objective Realities} 

 \title{A Sharper Image: The Quest of Science and Recursive Production of Objective Realities} 

% \title{A Sharper Image: The Scientific Quest and  \\ 	Recursive Production of Objective Realities} 

% \title{A Sharper Image: The Game of Science and \\ Recursive Production of Objective Realities} 
	
% \title{A Sharper Image: Playing the Game of Science \\ and Recursive Production of Objective Realities} 

\author{Julio Michael Stern\vspace{4mm}\footnote{   
 IME-USP – The Institute of Mathematics and Statistics 
of the University of S\~{a}o Paulo.  
 Rua do Mat\~{a}o, 1010, 05508-900, S\~{a}o Paulo, Brazil.   
e-mail: \textit{jstern@ime.usp.br}\, or\,      
	\textit{jmstern@hotmail.com}    
							}%\footnote  
							}%author  

\lhead{J.M. Stern}
\rhead{A Sharper Image} 

% \date{\today}    
 \date{\normalsize Published\,at: \textit{Principia: an international journal of epistemology}, 
 24,\,2,\,255-297,\,2020;  doi:10.5007/1808-1711.2020v24n2p255\,; \  
 Updated at \textit{arXiv} on April 21, 2022. \\ }   
   	
\maketitle 

	\mbox{} \vspace{-2mm} \mbox{}

\begin{abstract} 
 %%% 	
This article explores the metaphor of Science as provider of sharp images of our environment, using the epistemological framework of Objective Cognitive Constructivism. 
These sharp images are conveyed by precise scientific hypotheses that, in turn, are encoded by mathematical equations.   
Furthermore, this article describes how such knowledge is produced by a cyclic and recursive development, perfection and reinforcement process, leading to the emergence of eigen-solutions characterized by the four essential properties of precision, stability, separability and composability. 
Finally, this article discusses the role played by ontology and metaphysics in the scientific production process, and in which 
sense the resulting knowledge can be considered objective. \\ 
\textit{Keywords}: Objective cognitive constructivism; Eigen-solutions; Ontology; Metaphysics.  
 %%%      	
\end{abstract}  

 \mbox{} \vspace{0mm} \mbox{}

\begin{flushright} 
	
 \begin{minipage}{4.75in}   %{5.4in} 
 %\begin{justify} 
		  
 %\textit{ \mbox{} \hspace{-4.5mm}  
 %La vie est fascinante: il faut seulement %la regarder avec les bonnes lunettes}. \\  
 %Life is fascinating: one just has  to look at it using the right lenses. \\  %glasses
 %Alexandre Dumas fils (1824--1895). \\ 
 %\vspace{0mm} 

 \textit{ \mbox{} \hspace{-4.5mm}  
 You can dance in a hurricane;  \\ 
 But only if you are standing in the eye}.  
 \vspace{1mm} \\  
 The Eye (2015), by Brandi Carlile. \\  
 %\vspace{0mm} 

 \textit{ \mbox{} \hspace{-4.5mm}  
 O diabo existe e n\~{a}o existe?  
 Dou o dito. Abren\'{u}ncio. \\ % Essas melancolias.
 O senhor v\^{e}: existe cachoeira; e pois?  
 Mas cachoeira \'{e} barranco %\\ 
 de ch\~{a}o, e \'{a}gua se caindo por ele, retombando; o senhor
 consome %\\ 
 essa \'{a}gua, ou desfaz o barranco, sobra cachoeira
 alguma?  \\ 
 % Viver \'{e} negócio muito perigoso... \\ 
 Solto por si, cidad\~{a}o, \'{e} que n\~{a}o tem diabo nenhum. Nenhum! \\ 
 O diabo [existe] na rua, no meio do redemunho [redemoinho]}.  \vspace{1mm} \\  
 The devil exists and it doesn't? 
 %So I say it; 
 That is the message; isn't it? \\ 
 You see: a waterfall exists, right?    
 But a waterfall  is a steep step on \\ 
 the ground and water falling through it, tumbling down; you either \\ 
 consume the water or smooth the step, is there still a waterfall left?  \\ 
 %level the slope  
 % Living is a very perilous affair... \\ %business... 
 %By itself [als ding an sich] 
 %Standing all by itself, just loose,  
 %Standing alone, all by itself, 
 Cut loose, just by itself, 
 there is no such thing as a devil; none!  \\ 
 %there is no devil; none!   \\  
 The devil [exists] on the road, at the middle of the whirlwind.  \vspace{1mm} \\  
  Grande Sert\~{a}o Veredas (1956), by Jo\~{a}o Guimar\~{a}es Rosa. \\ 
 %\vspace{0mm}  
   		
 \end{minipage} 
	
\end{flushright}

 \mbox{} \vspace{1mm} \mbox{}

\section{The Sharper Image Metaphor}

\begin{figure}[tb] 
	%\mbox{} \vspace{0mm} \mbox{}    
	
\centerline{ 
	\includegraphics[height=1.8in]{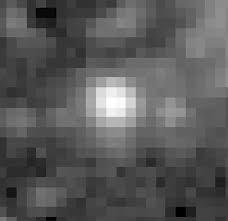} 
	\mbox{} 
	\includegraphics[height=1.8in]{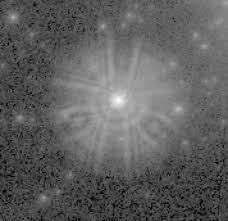} 
	\mbox{} 
	\includegraphics[height=1.8in]{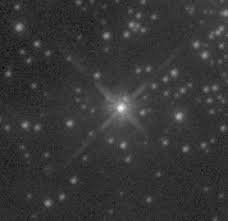} 
}%centerline 

\mbox{} \vspace{-0mm} \mbox{}  

\centerline{ %trim=l b r t 	
	\includegraphics[height=1.0in, 
	    trim= 169mm 0mm 169mm 35mm , clip]{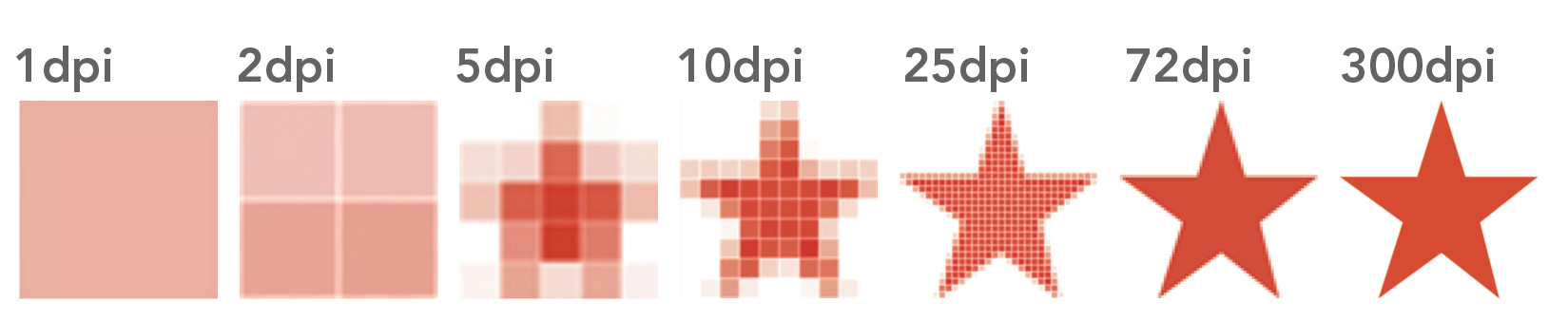} 
     \mbox{} \hspace{5mm} \mbox{}  
    \includegraphics[height=1.0in, 
    trim= 0mm 16mm 0mm 0mm , clip]{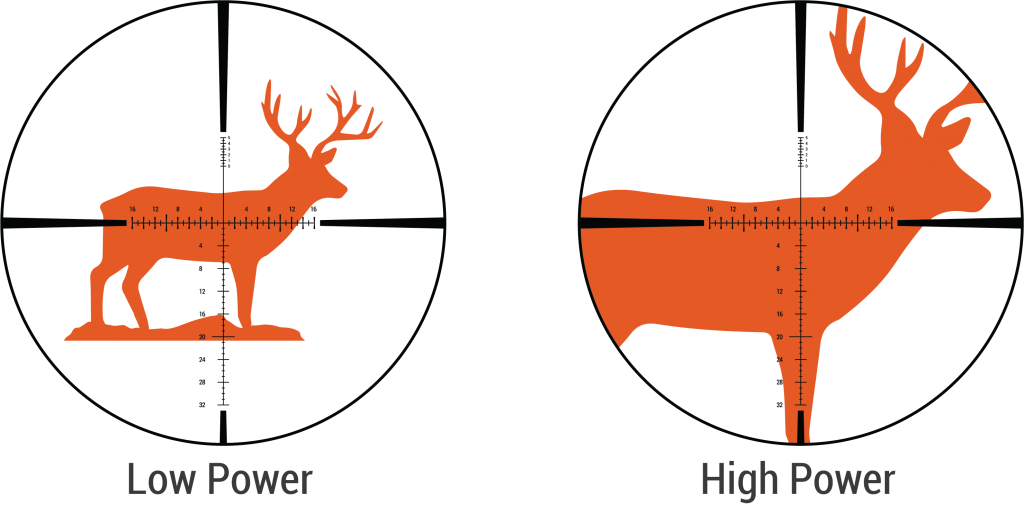} 
}     
 
\caption[]{(top) Blurry vs. Sharper (less aberration) images of star Melnick-34; \\ 
(bottom) Resolutions of 5, 10 \& 25 pixels/line; Magnification factors of 1x \& 2x.}    
	     
\end{figure} 

 Telescopes work by producing images of distant 
objects at higher magnification and better resolution 
than the observer's naked eye is capable of. 
The magnification factor (see Figure 1, bottom right) specifies how many times larger the observer sees an object, while resolution (see Figure 1, bottom left) specifies the observer's ability to distinguish apart (resolve) two close objects, that is, resolution refers to the image's sharpness or amount of fine detail it makes available to the observer.    
Figure 1 (top) displays three images of the same region of space, centered at Melnick-34 -- a distant star in the Tarantula Nebula.
These images were produced by:   
(a) The ground based telescope of the European Southern 
Observatory; The Hubble's wide field planetary camera, 
(b) before and (c) after the space shuttle Endeavour 1993 mission to correct a spherical aberration problem.   
While these images have the same magnification factor, their resolution  range from low(er) to high(er), from left to right, that is,  
images to the left are impaired by stronger aberration or distortion effects that make them more confused, blurry or indistinct, when compared to the sharper 
image to the right; 
Moreover the same images are impaired by the occurrence of artifacts -- spurious effects like pixelation, replications of bright points by dimmer copies around it, or halo-like effects around larger sources of light.

This article explores the metaphor of Science as provider of ``sharp images'' of our environment. 
As in the telescope example, the images we can possibly see obviously depend on the instruments we have the ability to build and, perhaps less obviously, also depend on the characteristics of our inborn (biological) sense of vision.   
Similarly, we argue that our ability to interpret and recognize those images is constrained by our cognitive capabilities, by our theoretical frameworks, etc.   
Accordingly, such intellectual abilities are considered essential for our ``mind's eye'' in producing and processing images that usefully  represent our environment.   
In this sense, the epistemological framework developed in this article belongs to the philosophical tradition of cognitive constructivism, as developed by 
Maturana and Varela (1980), Foerster (2003), Foerster and Segal (2001), Krohn and K\"{u}ppers (1990), Zeleny (1980)  and many others.       

Nevertheless, the epistemological framework developed in this article has a distinctly ``objective'' character that sets it apart from many alternatives in the constructivist tradition. 
Using our optical analogy once more, the quality of the images provided by a telescope can be characterized by tailor-made measures used to quantify magnification, resolution, and  specific aberration effects.  
Similarly, we claim that the quality of scientific representations can be quantitatively accessed and 
correctly measured. 
Such measures are the technical touch-stones needed to lay down the statistical, mathematical and logical foundations 
of the \textit{Objective Cognitive Constructivism} epistemological framework.  

In previous works we define a statistical measure tailor-made for the aforementioned purpose, namely, $\ev(H\g X)$,  the $e$-value, or {epistemic value of  hypothesis $H$ given observational data $X$}. 
The statistical, mathematical and logical properties of the $e$-value are perfectly adapted to support and to work in tandem with the 
\textit{Objective Cognitive Constructivism} 
epistemological framework, as further discussed in Sections 4 and 7.     
Nevertheless, this article does not focus on 
formal analyses of such mathematical constructs.  
Rather, its goal is to develop the optical metaphor  introduced in this section in order to explain, in an easy and intuitive way, some of the basic ideas and key insights used in this framework.  
In accordance with this goal, epistemological arguments are supported by figures illustrating analogies or providing visual context.  
In order to better develop our arguments we use,  as historical background, the science of optics as seen trough the works of  
Giambattista della Porta (1535-1615),  
Galileo Galilei (1564-1642), 
Johannes Kepler (1571-1630), 
Ren\'{e} Descartes (1596-1650), and 
Pierre de Fermat (1607-1665).

\section{Galileo's telescopes and astronomy, della Porta's natural magic, and Kepler's optical theory}

\begin{figure}[tb] 
	%\mbox{} \vspace{0mm} \mbox{}    
   
\centerline{ 
	\includegraphics[width=2.7in, %trim=l b r t  
	trim= 0mm 0mm 0mm 0mm , clip]{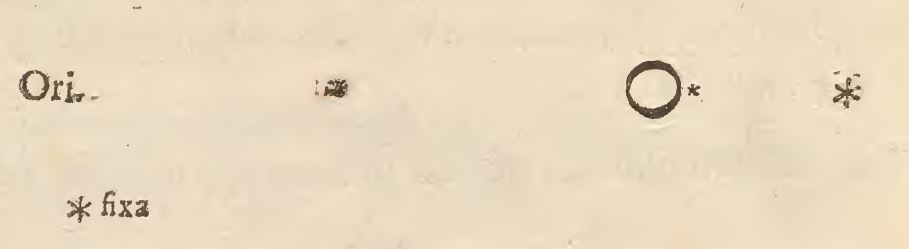}
	\mbox{} 
	\includegraphics[width=2.7in, %trim=l b r t  
	trim= 0mm 2mm 0mm 2mm , clip]{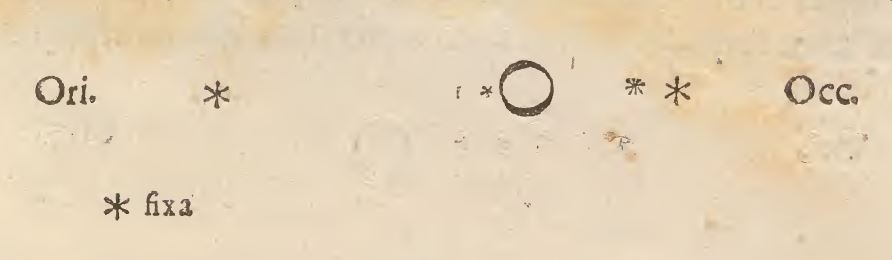} 
}%centerline 

\caption{Galileo (1610, p.26-27) reported observation of Jupiter's moon transit.} 
%Die.27 Ho.1 min.4 
 
\end{figure}

In 1610 Galileo Galilei reported the existence of four moons around planet Jupiter, a surprising announcement 
since, at that time, standard astronomical knowledge assumed that planets have no moons.  
Figure 2 depicts drawings by Galileo showing a fixed star and the four moons he could observe swinging around Jupiter.  
On one hand, this discovery should be welcome, for it could have very useful  practical applications. 
For example, moons's transit times could be computed and tabulated, serving as the basis for a clock available worldwide that, in turn, could serve as a much needed navigation tool used for determining the longitude at a given location, see Aksnes (2010). 
On the other hand, the addition of moons to the known planets was incompatible with the Catholic official doctrine of the time. 
Hence, it is not surprising that Galileo's discovery became the topic of heated debates. 

Galileo made his observations using a telescope, a recent Dutch invention that Galileo perfected and brought to Rome,  
and Jupiter's moons could only be observed with the aid of a such an instrument.   
Hence, a central topic of those debates: 
Should Jupier's moons, as seen through a telescope, be considered as  ``real'' or legitimate objects, or should they be dismissed as invalid artifacts or optical illusions?   
 
In a letter written in the same year of 1610 to Johannes Kepler, Galileo bitterly complains about his critics, as stated in the next quotation,  
see Santillana (1978, p.9). 
 % and Morinelli (2016, p.16).  
 % Karl Von Gebler, Galileo Galilei, p.26 (1879).  
  
\begin{quotation}  
\textit{My dear Kepler, what would you say of the learned here, who, replete with the pertinacity of the asp, have steadfastly refused to cast a glance through the telescope? What shall we make of this? Shall we laugh or shall we cry? %... \\   
Verily, just as serpents close their ears, so do these men close their eyes to the light of truth.}\footnote{Volo, mi Keplere, ut rideamus insignem vulgi stultitiam. Quid dices de primariis huius Gimnasii philosophis, qui, aspidis pertinacia repleti, nunquam, licet me ultro dedita opera millies offerente, nec Planetas, nec [Lunam], nec perspicillium, videre voluerunt? Verum ut ille aures, sic isti oculos, contra veritatis lucem obturarunt.
Letter of 19/08/1610 from Galileo to Kepler (Opere, v.X, Letter 379, p.421-423).}   
\end{quotation}

Galileo was prosecuted and judged by the Catholic  inquisition, ending his days in house arrest, see Finocchiaro (1989) for a detailed account.  
Over time, the ``Galileo affair'' backfired badly for the church, making Galileo a martyr of scientific progress in public opinion;  see for example the play Life of Galileo by Bertolt Brecht (1955). 
Hence, retrospectively, Galileo critics are frequently portrayed as obstinate fools that deny obvious conclusions or even refuse to see what is clearly put in front of their eyes, as Galileo complains in his letter to Kepler.   
Notwithstanding the scientific merits of Galileo, for the purpose of this article, it is important to carefully examine some arguments presented by his  critics, as well as some  counter-arguments presented by his contemporary supporters; this is the main objective of this section.

Cesare Cremonini (1550-1631) was a professor of Aristotelian philosophy at the university of Padua and a personal acquaintance of Galileo. 
Nevertheless, he vehemently refused to acknowledge the objects Galileo insisted could be seen trough the telescope, as attested in quotation in a letter written in 1611 by Paolo Gualdo (1553-1621) to Galileo: 

\begin{quotation}
\textit{[Cremonini:] I do not wish to approve of claims about which I do not have any knowledge, and about things which I have not seen; moreover, looking through those glasses confuses my head. 
Enough! I do not want to hear anything more about this.} 
 Galileo Opere vol.XI, no.564, p.165.\footnote{
  Questo \`{e} quello, dico, c'ha dispiacciuto al S.r Galilei, ch'ella non habbia voluto vederle. Rispose: Credo che altri che lui non l'habbia veduto; e poi quel mirare per quegli occhiali m'imbalordiscon la testa: basta, non ne voglio saper altro.  
  Letter of 29/07/1611 from P.Gualdo to Galileo; OG v.XI, doc.564, p.139-140.}   
\end{quotation}

Cremonini's dizziness or mind confusion (m'imbalordiscon la testa) when looking through Galileo's telescope can be interpreted as a reference to the blurry images produced by those primitive telescopes. 
Following the analysis of Bernieri (2012), the aberration effects in Galileo's 1610 telescopes should impose a resolution limit of two and a half Jupiter's equatorial radii for resolving a moon from the planet's edge. 
At Figure 2, Galileo seems to report a situation beyond what he could directly observe. 
Our point is not to question Galileo's drawings or data, but to highlight the importance of taking into account the quality of Galileo's instruments, measured by how sharp or blurry were the images they could provide.  
After all, it is easy to understand how low quality telescopes could undermine the credibility of Galileo's observations --  
it would be difficult for Galileo to convince someone by asking him or her to look at something he or she could not directly and clearly see.  
Cremonini denial of the new ``objects'' presented by Galileo's telescope are further analyzed in Muir (2007),  Boorstin (1983) and Biagioli (1993, p.238). 
Fortunately, Galileo's discoveries triggered a rapid development of lens grinding techniques and telescope construction technology, improving image magnification and resolution, and dissipating concerns in the line of those raised by Cremonini.

Christopher Clavius (1538-1612), an influential  Jesuit mathematician and astronomer, rejected Galileo's discoveries based on rather different concerns. 
He argued that telescopes are instruments designed with the specific purpose of creating optical illusions. 
Hence, any conclusions based on telescope observations should be, in principle,  dismissed as ``fake news''.

\begin{quotation}
\textit{Clavius said  ...about the four stars [moons of Jupiter], 
 that's ridiculous, for one needs to manufacture a spyglass that creates them for next display them, and that  Galileo can have his opinion and he shall have his own.} 
Galileo Opere, vol.X, p.442).\footnote{[Clavio] disse   
	...delle quattro stelle, che se ne rideva, 
	et che bisognier\`{a} fare uno ochiale che le 
	faccia e poi le mostri, et che il Galileo tengha 
	la sua oppinione et che egli terr\`{a} la sua.} 
\end{quotation}

\begin{figure}[tb] 
	%\mbox{} \vspace{0mm} \mbox{}    
	
	\centerline{ 
		\includegraphics[height=2.0in]{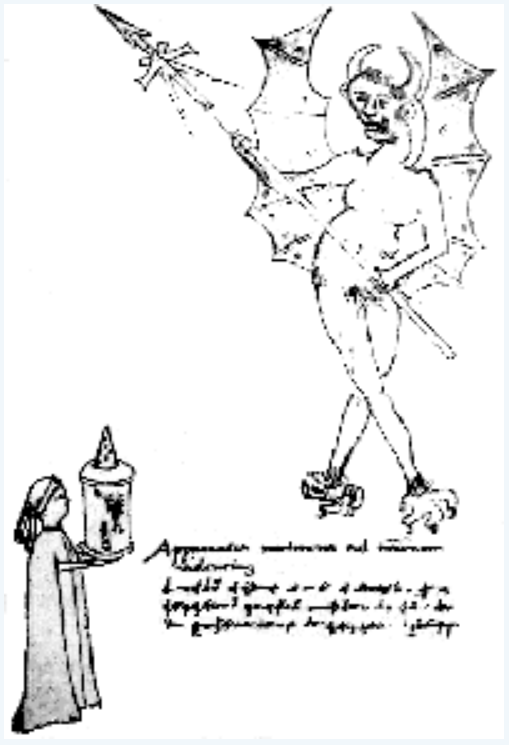} %2jpg 6png 8jpg
		\mbox{} \ \ \mbox{} 
		\includegraphics[height=2.0in, width=1.5in,  %trim=l b r t  
		trim= 0mm 20mm 15mm 5mm , clip]{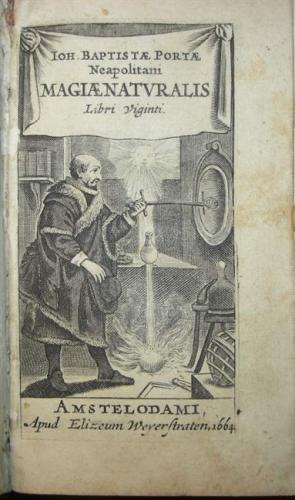} %4color 5BW 
		\mbox{} \ \ \mbox{} 
		\includegraphics[height=2.0in,  width=1.3in]{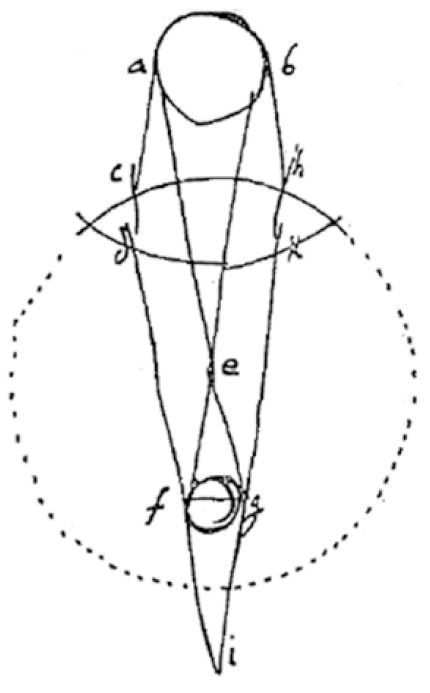} %2jpg 6png 8jpg 
		%\mbox{} 
		%\includegraphics[height=1.7in]{Lamp01.jpg} 
	}%centerline 
	
	\caption{Optics books by Fontana (1420) and  
		Porta (1589, 1658 \& 1610). } 
	% Gebr\"{u}der Bing N\"{u}rnberg, Germany (c.1900) 
	
\end{figure}

At the time of Galileo's discovery, most available books treating the area of optics, like Fontana (1420) and Porta (1558), teached the art of \textit{Magia Naturalis}, namely, how to perform  wonderful tricks and surprising experiments that could be produced by natural means and used to  entertain the public, including burning glasses, mirror chambers and magic lanterns projecting illusions, see Figure 3 (left and center).   
Considered in the context of this cultural background, Clavius' objections do not seem unreasonable.

In May 1611, Luca Valerio (1553-1618), a Jesuit mathematician, formulated an epistemological counter-argument to the aforementioned objections, based on  evidence provided by repeated and consecutive observations, see next quotation and also Biagioli (2006, p.111) and Favaro (1983, v.I, p.573).   
On one hand, pragmatically, Valerio's argument could justify the use of methods devised to filter out common artifacts from crude observational data, correct regular distortions, or interpolate between observations, as routinely done in modern astronomy. 
On the other hand, epistemologically, Valerio's argument provides some support for trusting a telescope, even lacking any explanation of how and why such an instrument works. 

\begin{quotation} 
	\textit{It has never crossed my mind that the same glass [always] aimed in the same fashion toward the same star [Jupiter] could make it appear in the same place, surrounded by four stars that always accompany it ... in a fashion that one evening they might appear, as I have seen them, three to the west and one to the east [of Jupiter], and other times in very different positions, because the principles of logic do not allow for a specific, finite cause [the telescope] to produce different effects when [the cause] does not change but remains the same and maintains the same location and orientation}. 
\end{quotation}

In his later works, della Porta (1610) tries to  explain how optical instruments work. Nevertheless, despite their diagrammatic or geometric appearance, these explanations are mostly descriptive and qualitative in nature, see Figure 3 (right), Heeffer (2017) and Borrelli (2017).   
Galileo himself offered no explanations on the inner workings of a telescope; 
on the contrary, he protected  the technology of telescope design and construction as trade secrets, for selling telescopes, or giving them as gifts to influential people, was for him a source of income and prestige, see Biagioli (2006, p.127).

In the same year of 1610, Kepler answered Galileo's letter, expressing a very different attitude; for an inspiring short biography see Koestler (1960), 
for Kepler's general attitude concerning theoretical explanations see Holton (1988), Martens (2000) and Stephenson (1987).    
For Kepler, providing a good optical theory and explaining the bases of telescope technology was an important task to perform in gaining the public trust, as stated in the next quotation.

\begin{figure}[tb] 
	%\mbox{} \vspace{0mm} \mbox{}    
	
	\centerline{ 
		\includegraphics[height=1.5in]{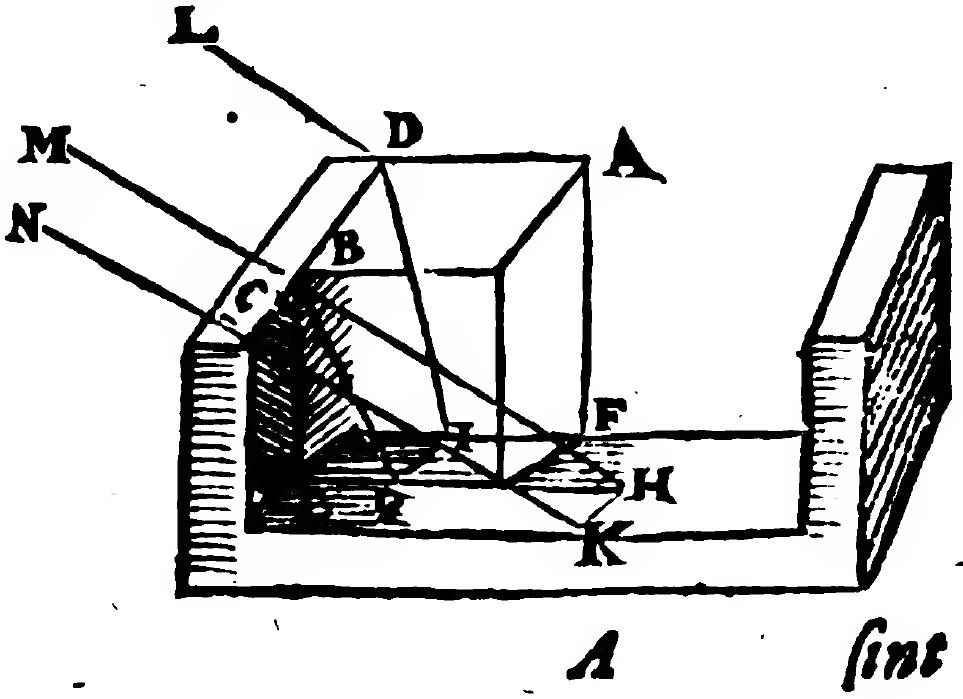}
		\mbox{} \hspace{3mm} \mbox{} 
		\includegraphics[height=1.6in, width=1.0in, angle=0]{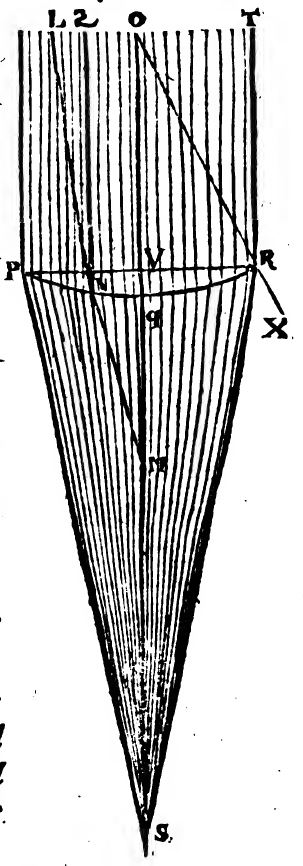}
		\mbox{} \hspace{3mm} \mbox{} 
		\includegraphics[height=1.4in]{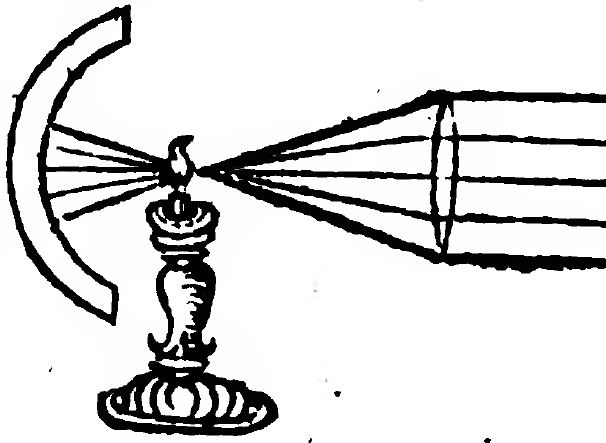}
	}%centerline 
	
	\mbox{} 
	
	\centerline{ 
		\includegraphics[width=5.5in,, angle=0]{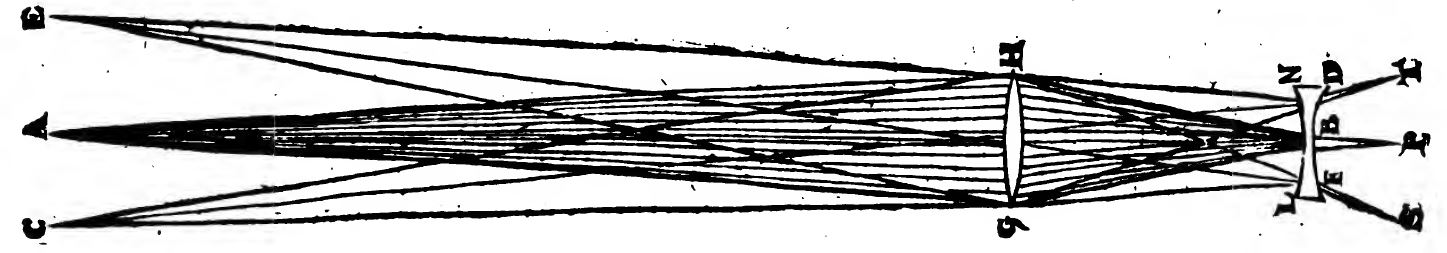}  % width=0.8in,
	}%centerline
	
	\mbox{} \vspace{1mm} \mbox{} 
	
	\centerline{ 
		\includegraphics[width=5.5in,, angle=0]{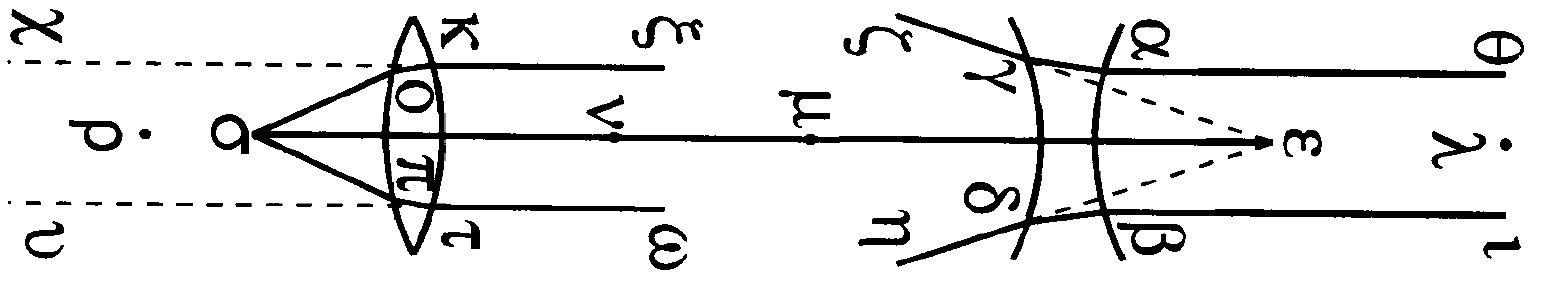}  % width=0.8in,
	}%centerline %KeplerDonahue.p.218 pdf.117 
	
	%\mbox{} \vspace{1mm} \mbox{} 
	
	\caption{  
		Kepler diagrams explaining his refraction law (top left), simple optical 
		%\linebreak  
		elements like lenses and mirrors (top right), and compound instruments (bottom).} 
	
\end{figure}

\begin{quotation}  
	\textit{In the Optical part of astronomy, I gave and 
		explained a lucid geometrical demonstration   
		of what happens in simple lenses.   
		%[120] p.17    
		...moreover, concerning    %and likewise, 
		the widely available circular tubes with two lenses, 
		including the very instruments you, Galileo, 
		have used to pierce the heaven:  
		I am trying to convince the incredulous to have 
		faith in your instruments.}   
	%[127] p.18 incredulis 
	Kepler (1610, p.17-18), 
	Dissertatio cum Nuncio Sidereo.\footnote{
		In astronomiae parte Optica ut quid in simplicibus 
		perspicillis accideret luculenta demonstratione 
		geometrica redderem expeditum.  %CSN p.17 fim  
		...adeoque \& inter ipsos vulgo circulatos 
		tubos bilentes, \& inter tuam Galilaee machinam, 
		qua coelum ipsum terebrasti:\,sed nitor hic fidem 
		incredulis facere instrumenti tui.}  %CSN p.18 fim 	
\end{quotation}

Figure 4, from Kepler's \textit{Astronomiae pars Optica}, shows some diagrams (original and modern versions) used by Kepler to explain his optical theory, see Kepler (1604; 1611; 2000, p.218), Heeffer (2017), Malet (2003) and Molesini (2011).        
Appendix A presents this theory, known today as \textit{paraxial optics}, in a very simple and compact form with the use of the modern language of Linear Algebra. 

Kepler's theory follows a three-step program in the task of explaining how and why a telescope works, steps that could be characterized as follows: 
\begin{description} 
	
	\item[Scientific laws:] Provide a sound theoretical framework based on laws formulated as \textit{mathematical equations}. In the case at hand, the pertinent scientific law relates the incidence and refraction angles of a light ray crossing the boundary between two propagation media, like from air to glass or vice-versa.  
	
	\item[Analysis:] Identify key elemental parts of a system and provide a good understanding of how and why such \textit{single or separate elements} work. 
	In the case of a refracting telescope, the key optical elements are (thin, spherical, glass) lenses. 
	
	\item[Synthesis:] Provide \textit{compositionality rules}, teaching how to put together several single elements in order to construct a complex system. 
	In the case of a refracting telescope, this comprises means and methods for specifying  distinct lenses and their location in a tubular supporting structure in order to produce sharp images.        	
	
\end{description} 	

The next sections will examine and further explore Kepler's three-step program. Section 3 examines this program as a didactic approach to teaching science. 
Section 4 explores the epistemological consequences of  this program and the fundamental role it plays in the Objective Cognitive Constructivism epistemological framework.

\section{ Playing the Game of Science:   \\ 
	      Objective Cognitive Constructivism}

 \textit{Os Cientistas} (The Scientists) series 
 was a joint project launched in 1972 by FUNBEC - 
 the Brazilian Foundation for Science Teaching, 
 %Funda\c{c}\~{a}o Brasileira para o Ensino da Ci\^{e}ncia 
 and Abril Cultural, a publisher specialized in magazines and inexpensive books, see Pereira (2005).    
 The series had 50 issues in total, sold every 
 2 weeks  at news stands nation wide. 
 Each issue contained: 
 (a) A fascicle with a richly illustrated and 
 historically situated biography of an important scientist;  
 (b) An instruction manual for doing empirical experiments related to that scientist's work; and 
 (c) A small box containing necessary materials. 
 The cost of each issue was equivalent to only a couple of US Dollars, making it easily accessible and affordable to teenagers interested in learning science in Brazil, at that time a fairly low-income country, 
 see Civita (1972) and Pereira (2005).   
 A didactic and commercial success,   
 %(first edition sold around  200,000 copies),    
 the series was re-edited a few times during the 
 70's and 80's, until it was discontinued due to 
 new safety standards in the toy industry.

\begin{figure}[tb] 
	%\mbox{} \vspace{0mm} \mbox{}    
    
 \centerline{  
 	\includegraphics[height=2.2in, %trim=l b r t  
 	trim= 0mm 0mm 0mm 0mm , clip]{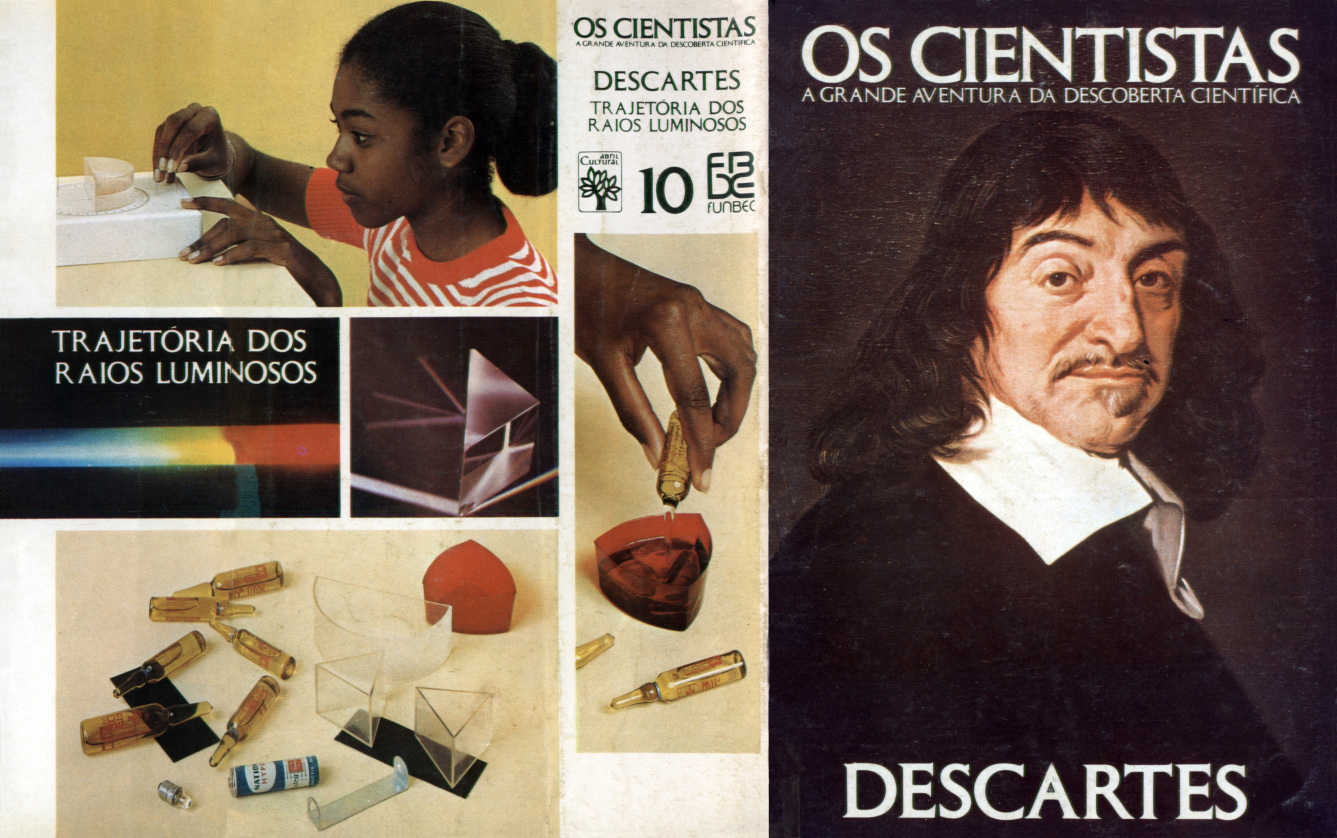} 
 	\mbox{} 
 	\includegraphics[height=2.2in, %trim=l b r t  
 	trim= 0mm 0mm 0mm 0mm , clip]{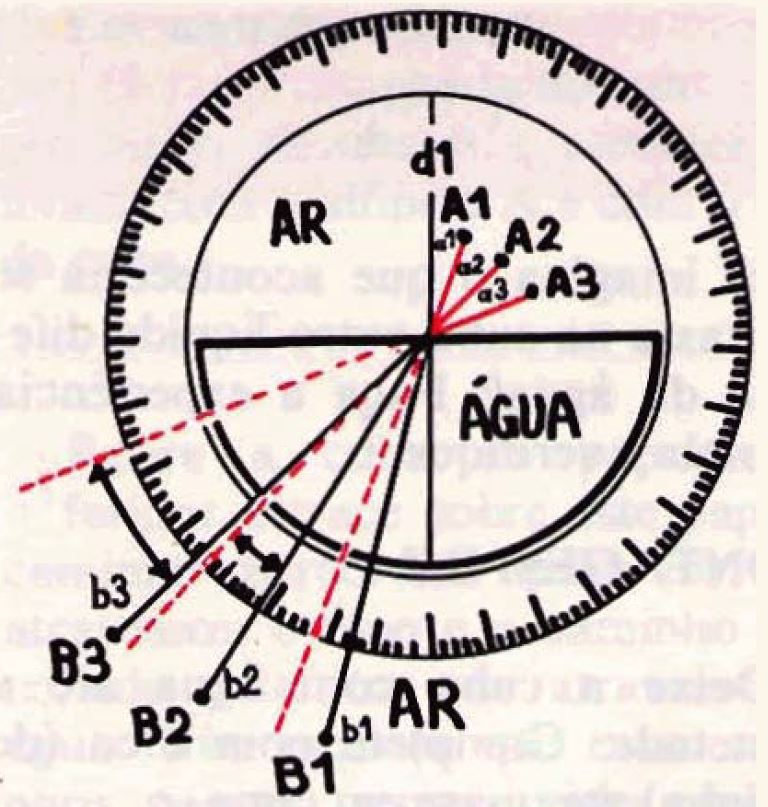} 
 }%centerline 

 \caption{The Scientists  
 	- Descartes - kit of experiments.} 
    
\end{figure}

 \textit{Os Cientistas} issue featuring Ren\'{e} 
 Descartes, suggests several experiments for 
 understanding his refraction law. 
 The provided materials include a battery lantern 
 as light source, slit-screens as simple collimators,   
 printed scales used for angular measurements, 
 and a few small acrylic containers shaped 
 as a half-cylinder, prisms, etc, see Figure 5.  
 These containers can be filled with  liquids 
 of different refractive indexes (approx.value), 
 like water (1.33), olive oil (1.45), 
 monochlorobenzene (1.53), etc. 
 %tetrachlorobenzene (1.8), etc.  

\begin{figure}[tb] 
	%\mbox{} \vspace{0mm} \mbox{}    
 
 \centerline{ 
 	\includegraphics[height=1.7in, %trim=l b r t  
 	trim= 0mm 0mm 0mm 0mm, clip]{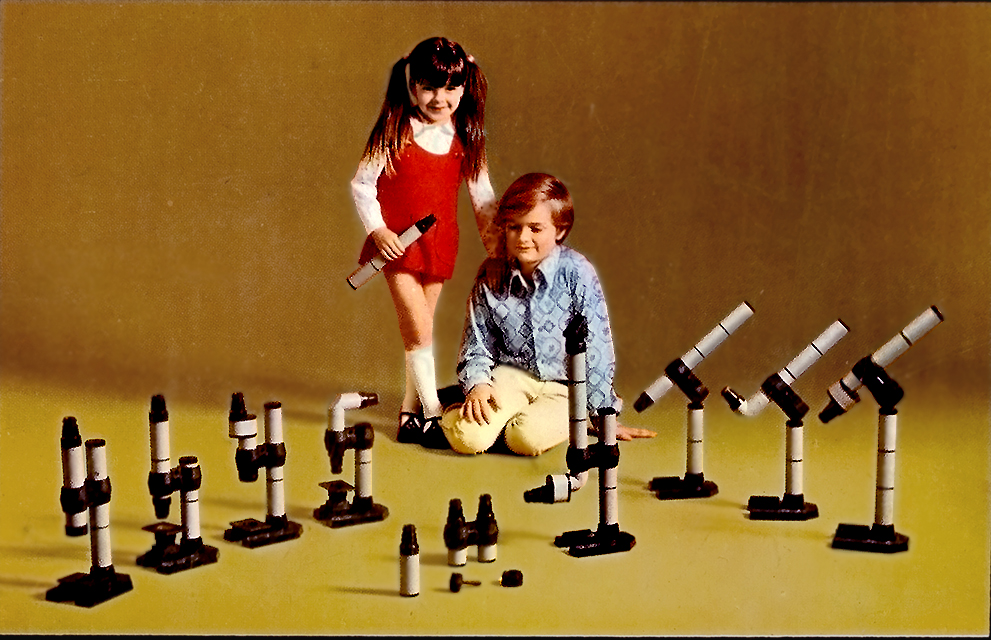} 
 	% trim= 0mm 0mm 0mm 0mm, clip]{Optics90.png} 
 	\mbox{} 
 	\includegraphics[height=1.7in, %trim=l b r t  
 	trim= 0mm 0mm 0mm 85mm, clip]{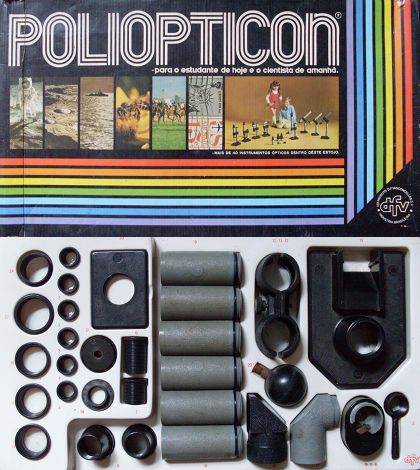}  
 }%centerline

 \caption{Polyopticon - Optical components and construction kit} 
     
\end{figure}  
 
 Polyopticon, manufactured by the Brazilian company 
 D.F.Vasconcellos, was a ``construction kit'' for optics, see Figure 6.  
 It was sold at toy stores, remained in production from 
 around 1960 to 1980, and was also exported to the USA.  
 The kit contains a variety of simple optical elements, 
 like a prism, mirrors, and lenses of several shapes. 
 Each lens is mounted on a supporting ring, and  
 pieces of plastic tubing are used to connect 
 multiple elements. 
 Supporting rings and tubes are all screw-threaded, 
 allowing for easy but stable and precise adjustment of 
 distances between well-aligned optical elements. 
 The construction kit allows the assembly of 
 a variety of microscopes, telescopes, and other 
 optical instruments. 

\begin{figure}[tb] 
	%\mbox{} \vspace{0mm} \mbox{}     

\centerline{  
	\includegraphics[height=1.7in, %trim=l b r t  
	trim= 0mm 0mm 0mm 0mm , clip]{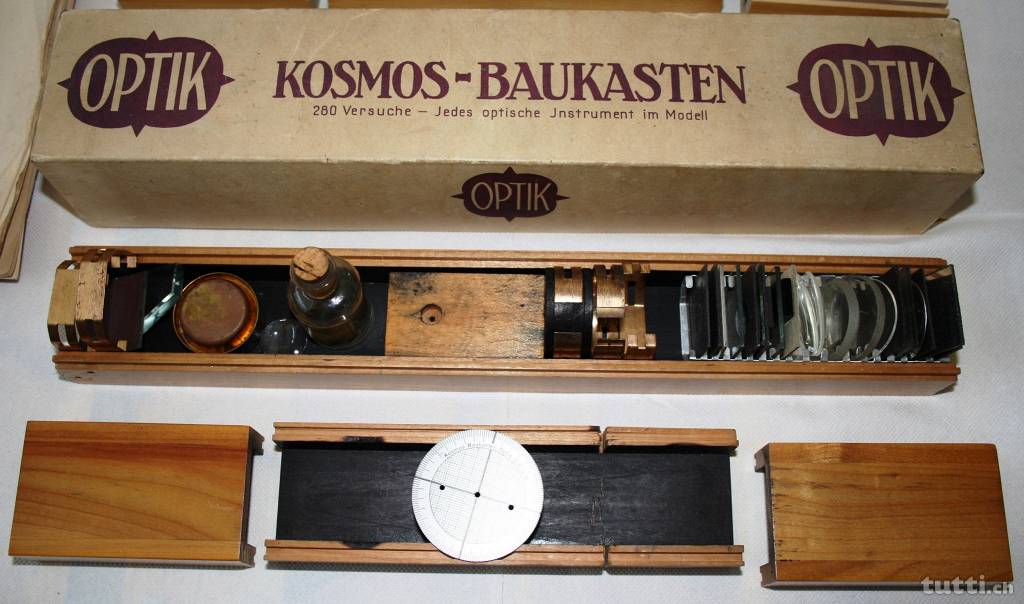} 
	\mbox{} %trim=l b r t  %b=62 
	\includegraphics[height=1.7in, width=2.5in,    
	trim= 0mm 0mm 0mm 0mm , clip]{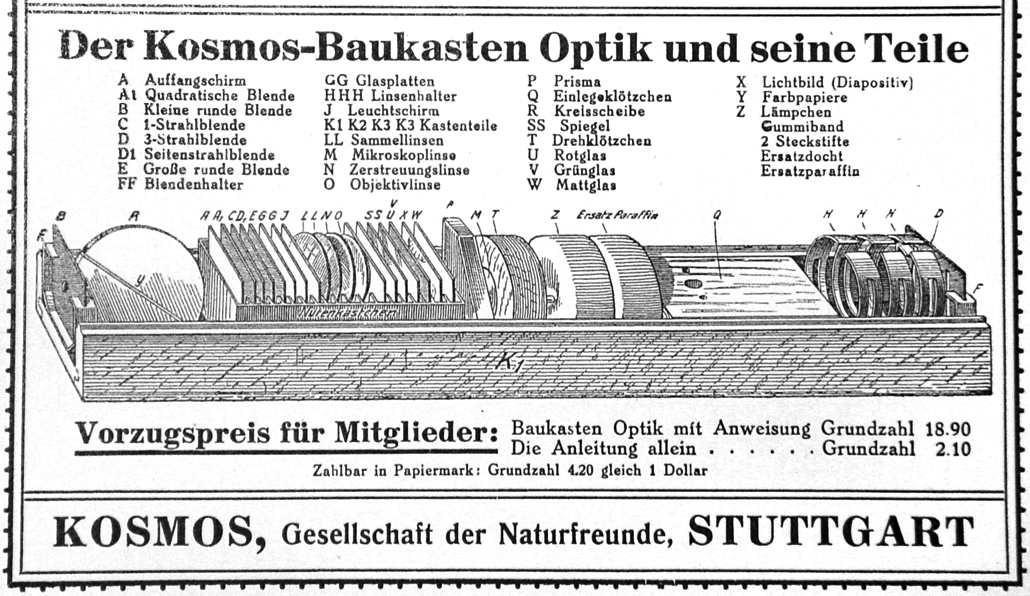} 
}%centerline 

\caption{Kosmos-Baukasten, Optik , 1923.}     
  
\end{figure} 

The preceding examples are late descendants of earlier efforts on popular scientific education;  
among those, some of the best known and most influential are the celebrated Kosmos Bauk\"{a}sten, manufactured in Germany from around 1920 to the present day, see Figure 7 and Fr\"{o}hlich (1923).    
All the aforementioned didactic resources are concerned with teaching of basic optics, and all of them seem to promote  
Kepler's approach for  ``guiding the perplexed'' 
into a path of understanding.   
Specifically, they focus on one or more of the  
following steps: \\ 
(a) Explain the \textit{precise and invariant laws} that govern  the trajectory of light rays when reflected or refracted at smooth interfaces between distinct propagation media;   \\ 
(b) Explain the behavior of \textit{simple} optical \textit{elements}, like mirrors, prisms or lenses;  \\  
(c) Explain how to \textit{compose} or assemble \textit{complex systems} putting 
together simple(r) optical components. \\   
This three-step approach is closely related to the cognitive constructivist fundamental insight on the construction of objective realities, to be presented at the next sections.

\begin{figure}[tb] 
	%\mbox{} \vspace{0mm} \mbox{}    
 
\centerline{  %trim=l b r t     	 
	\includegraphics[height=2.3in, %width=1.5in,    
	trim= 0mm 0mm 0mm 0mm , clip]{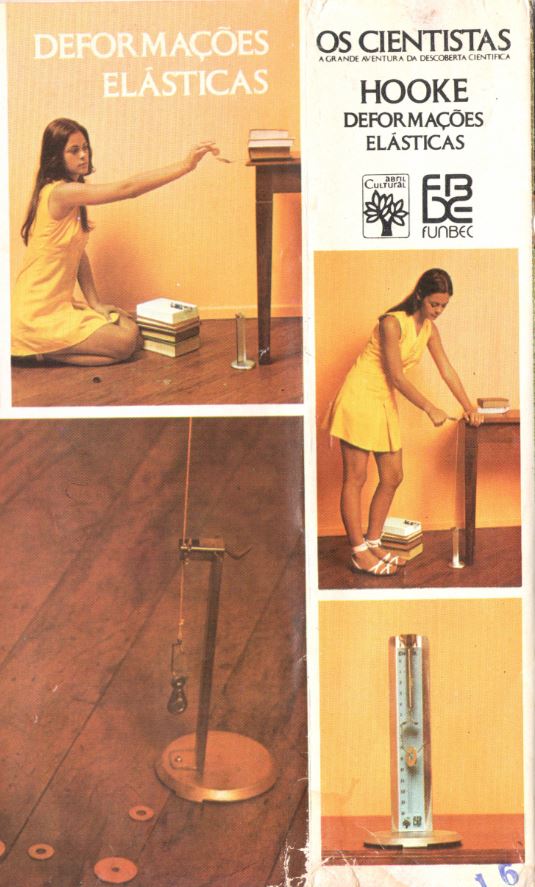}   %14
	\hspace{-1.8mm}  
	\includegraphics[height=2.3in, %width=0.78in,    
	trim= 0mm 0mm 0mm 0mm , clip]{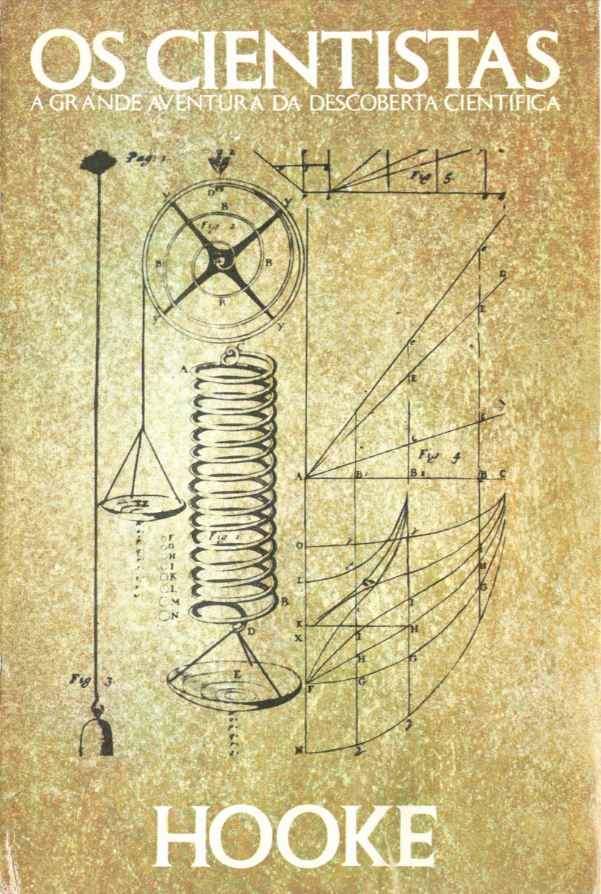}   %13
	\mbox{}  
	\includegraphics[height=2.3in, width=1.0in, %trim=l b r t  
	trim= 0mm 0mm 0mm 0mm , clip]{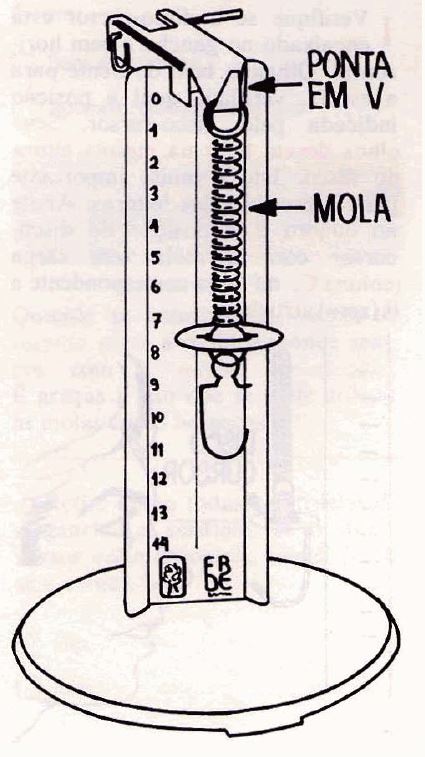}   	
	\mbox{}   	 
	\includegraphics[height=2.3in, width=1.2in,    
	trim= 0mm 0mm 0mm 0mm , clip]{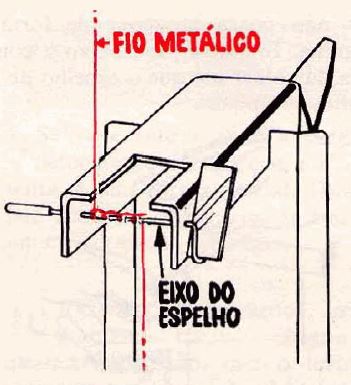}  
}%centerline %trim=l b r t  

  \mbox{} \vspace{3mm} \mbox{}     
  
\centerline{ 
	\includegraphics[height=1.2in, width=1.2in, angle=0,   
	trim= 0mm 0mm 0mm 0mm , clip]{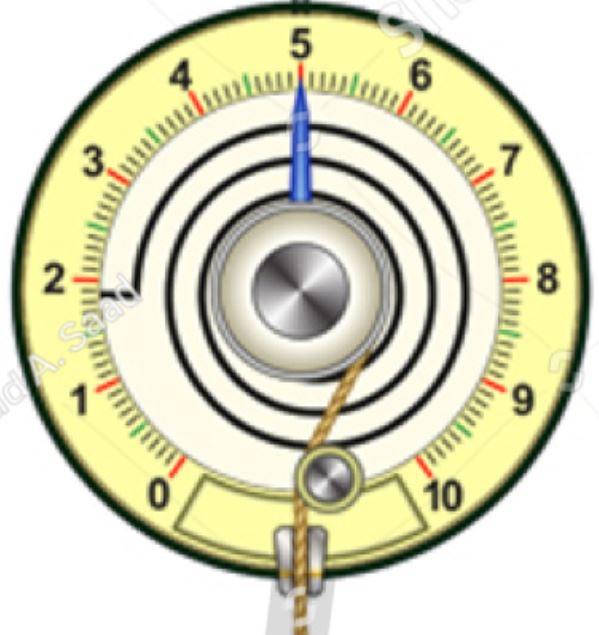} 
	\hspace{2mm}  %trim=l b r t  
	\includegraphics[height=1.3in, %width=1.7in, %trim=l b r t  
	trim= 0mm 0mm 0mm 0mm , clip]{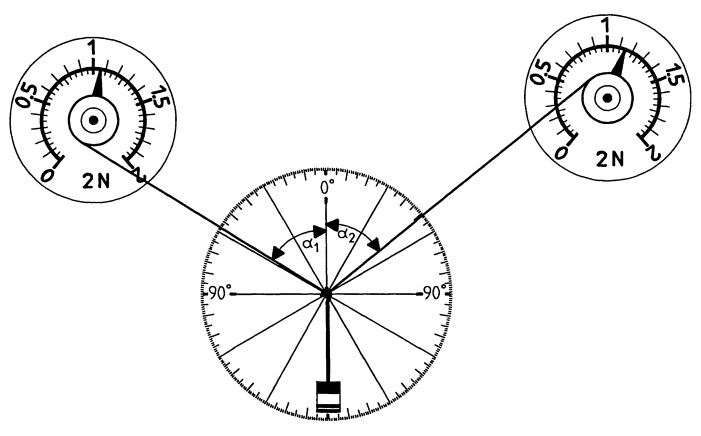} 
	\hspace{2mm} 
	\includegraphics[height=1.3in, width=1.5in, %trim=l b r t  
	trim= 0mm 0mm 0mm 5mm , clip]{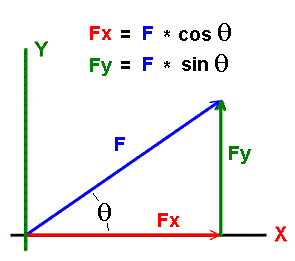} 
}%centerline 

  \caption{Hooke's law, force measure, and  (de)composition principles}       
 
\end{figure}

In the next sections, we will continue to explore the metaphor of science as provider of sharper images of our world and, therefore, will continue to stress optical analogies.     
Nevertheless, it should be clear that the epistemological framework of Objective Cognitive Constructivism applies to any discipline of exact sciences and, with the necessary adaptations, also far beyond. 
Figure 8 (top) shows ``Os Cientistas'' issue featuring  
Robert Hooke (1635-1703) and the accompanying  
experimental kit treating his deformation law, how to test it, and how to use it to build dynamometers for measuring forces. 
Figure 8 (bottom) depicts spiral-coil dynamometers, and how to use them to demonstrate and check the parallelogram law in the context of static force (de)composition;  
for a detailed discussion of the Objective Cognitive Constructivism epistemological framework in this application context, see Stern (2017a).

\section{Recursive Production of Objective Realities:  \\ 
Circularity and Emergence of Eigen-Solutions} 

%\section{Recursive Production of Objective Realities:  \\ 
%	Objects are Tokens for Eigen-Solutions}           

Galileo's telescopes did not appear out of nothing nor all at once.  
On the contrary, they have evolved along a well documented historical process over a period of several centuries. 
Ancient Greek and medieval Arab documents already describe the magnifying effect of spherical and lens shaped glasses, some of this knowledge being lost or forgotten during the European middle ages, see Bellosta (2002), Khan (2015), Mihas (2008) and Rashed (1993).   
European paintings and documents from the end of the middle ages attest the use of lens spectacles to remedy a variety of vision deficiencies, see Ilardi (2007).       
Fracastoro (1538, p.18), as quoted in Helden (1977, p,28), makes reference to double-lens spectacles used by visually impaired patients.  
Finally, in the turn of the XVI to the XVII century, Dutch and Italian lens grinders learned how build a simple telescope by installing two lenses into a tube,  
with either two convex lenses or one convex and one concave lens, see Helden (1977, p.15, 35, 50),  Ilardi (2007, p.209, 218), Porta (1589, 1658, p.368) and    
Sirtori (1618). 

From the Objective Cognitive Constructivism epistemological perspective, a lot can be learned by analyzing this historical evolution process:  
On one hand, it is important to realize that mankind did not have in advance a well determined road map to follow;  
on the contrary, free trial and random experimentation had an important role to play in this process.      
On the other hand, this process was not just driven by wild guessing and sheer luck;   
on the contrary, this process had a convergent or self-organizing  nature that requires close attention.     
Figure 9 gives a diagrammatic representation of several important aspects of this process and how they  interact:

\begin{figure}[tb]  
	\centering  
	\begin{small}  
		\begin{tabular}{c c c c c} 
			\cline{1-1} \cline{5-5}  
			Theoretical & \bvert & Metaphysical  & \bvert & Technological \\    
			\cline{3-3}  \\ 
			Mathematical  & $\Rightarrow$ & Causal  
			&  $\Rightarrow$ & Instrument   \\ 
			formalization   & & explanation   & &  specification   \\ 
			$\Uparrow$   & &               & & $\Downarrow$        \\  
			Speculative  &  \multicolumn{3}{c}{Sharp$^*$} & Project design   \\ 
			interpretation &  \multicolumn{3}{c}{images} &  \& calculation \\     
			$\Uparrow$   & &               & & $\Downarrow$         \\ 
			Aberration   & & Calibration/  & &  Fabrication    \\ 
			analysis & $\Leftarrow$ & observations    
			& $\Leftarrow$ &  /assembly \\  \\ 
			\cline{3-3}     
			%\multicolumn{2}{l|}{Sample space} & Operational & 
			%\multicolumn{2}{|r}{Parameter space} \\ 
			optics & \bvert & Operational & \bvert & 
			know-how \\ 
			\cline{1-1}  \cline{5-5}  \\ 
			\multicolumn{5}{c}{ {\bf Figure 9:} 
				Optical instrumentation production diagram.}
		\end{tabular} 

\mbox{} \vspace{15mm} \mbox{} 

		\begin{tabular}{c c c c c} 
	\cline{1-1} \cline{5-5}  
	Theoretical & \bvert & Metaphysical  & \bvert & Technological \\    
	\cline{3-3}  \\ 
	Mathematical  & $\Rightarrow$ & Causal  
	&  $\Rightarrow$ &  Hypotheses  \\ 
	laws and models   & & explanation   & &  formulation    \\ 
	$\Uparrow$   & &               & & $\Downarrow$        \\  
	Speculative  &  \multicolumn{3}{c}{Sharp$^*$/ precise} & Empirical  \\ 
	interpretation &  \multicolumn{3}{c}{laws and rules} & trial  design  \\     
	$\Uparrow$   & &               & & $\Downarrow$         \\ 
	Statistical   & & Data & &  Experiment    \\ 
	analysis & $\Leftarrow$ &  processing     
	& $\Leftarrow$ &  execution \\  \\ 	
	\cline{3-3}     
	%\multicolumn{2}{l|}{Sample space} & Operational & 
	%\multicolumn{2}{|r}{Parameter space} \\ 
	knowledge & \bvert & Operational & \bvert & 
	know-how \\ 
	\cline{1-1}  \cline{5-5}  \\ 
	\multicolumn{5}{c}{ {\bf Figure 10:} 
		Scientific production diagram.} 
\end{tabular} 
	\end{small} 
\end{figure} 
\addtocounter{figure}{2}

In the optical instrumentation production diagram,  
the right column represents technological know-how required to specify the components, like the shape of lenses and mirrors, and their correct assembly into an instrument, like a telescope. 
The bottom row of the diagram represents operational know-how, like lens or mirror grinding techniques, a business that requires specialized tools and materials. 
Moreover, an artisan manually grinding a component will hardly get it perfect at the first try; on the contrary this is a gradual process based on careful measurement of the component characteristics and shape corrections, see Section 6 for further details.  
In the same way, the assembly of a telescope requires gradual calibration of lens alignment and positioning in order to obtain the clearest images. 
The production and development of optical instruments also requires a circumstantial understanding of the aforementioned procedures, and such an understanding is best provided in the context of an appropriate theoretical framework, as already analyzed in sections 2 and 3; this is represented at the left column of the diagram.  
Understanding -- why things are or behave and why they do -- constitutes, by definition, metaphysical or causal explanations, represented at the top row of the diagram. 
Causal explanations are needed to make sense of what is routinely done and, every once in a while, can also provide innovative intuitions. 
    
Innovation can be beneficial all along the production cycle: New ideas can enhance the design and specification of instruments and components, refine production means and methods, or even improve the background theory itself. 
However, not every innovation proposal is successful -- in fact, only very few prevail. 
Therefore, it is important to have good criteria to distinguish what works from what does not.

The precise formulation of good performance criteria can become quite sophisticated and technical, see Section 6.  
Nevertheless, the essential motivation for such criteria is easy to comprehend: 
Pragmatically, good optical instruments should present sharp(er) images of our world. 
This pragmatic ideal is expressed in several versions of a well-known pun: \\ 
$\bullet$ The proof of the pudding (or putting) is in the tasting (or testing); \\ 
$\bullet$ \textit{La pruova del testo \`{e} la torta},  \\  
-- The test of the text (or baking pot) is the pie (or the twisted thing itself).   

As clearly displayed in Figure 9, the optical instrumentation production diagram has a cyclic configuration. 
All its components are circularly linked, and their development imply cross-referent and recursive processes. 
Finally, sharp images obtained from good optical instruments are represented as emerging entities at the center of the diagram. 
This representation conveys the idea that such fine images are the result of an essentially  recursive development and perfecting process.

Figure 10 depicts the Scientific production diagram as a close analog to the previously presented optical instrumentation production diagram. 
Instead of sharp images, precise scientific laws and exact compositional rules occupy the center of the diagram.   
In mathematical statistics, precise laws are represented by \textit{sharp statistical hypotheses} formulated as \textit{mathematical equations}, and these hypotheses can, in turn, be tested for their precision and accuracy.  
Technical aspects of design of empirical trials or experiments, data collecting and processing tools, and statistical analysis methods are represented at the right column and bottom row of the diagram. 
The use of mathematical models and development of scientific theories is represented at the left column of the diagram. 
Finally, causal explanations are represented at the top row of the diagram. Causal explanations are needed to make sense of routine scientific activities and, occasionally, can also provide innovative intuitions leading to new scientific hypotheses. 
Finally, precise scientific laws and exact (de)composition rules obtained from good scientific theories are represented as emerging entities at the center of the diagram. 
This representation conveys the idea that such fine scientific laws are the result of an essentially  recursive development and perfection process.

As it is typical of evolutionary processes, either in the case of optical instruments or in the case of scientific disciplines, development and perfection histories can \textit{not} be characterized, in the long run, by homogeneous convergence, but rather by a mixture of relatively smooth processes, like the fine-tuning of theory and experimental methods within a given framework, in combination with more abrupt jump processes or sudden transitions between distinct frameworks. 

For example, in the XVII century, reflecting telescopes were introduced as a novel technology offering significant advantages over refracting telescopes, see King (1955).  
As a second example, Venables (2017) shows how Amateur Telescope Making was made possible by two almost simultaneous breakthroughs by L\'{e}on Foucault (1858, 1859): First, the invention of a chemical process for silver coating glass; second, the invention of a new simple and efficient procedure for testing and correcting the shape of a mirror  during the grinding process.  
   
An example of discontinuous transition between theoretical frameworks is the replacement of Kepler's by Descartes' law of refraction, see Section 5 for further comments. 
Kepler's paraxial optics establishes a linear relation that can be stated (see Appendix A) as $n_1 \theta_1 = n_2 \theta_2$, where 
$\theta_1$ and $\theta_2$ are the incidence and refraction angles of a light ray crossing the boundary between propagation media characterized by refraction indices $n_1$ and $n_2$, receptively, see Figure 12 (right).    
In contrast, Descartes' law of refraction states a  non-linear relation between the aforementioned variables of interest, namely,    
$n_1 \sin \theta_1 = n_2 \sin \theta_2$. 
 %%% 
Descartes' law was a remarkable breakthrough, even considering that, for small angles (as is often the case for thin lenses used inside a telescope), Kepler's paraxial law provides a good approximation,  
 %%%
  $\theta << 1 \Rightarrow$  
  $\sin \theta \approx \tan \theta \approx \theta$.    
 %%% 
For further comments on  evolution of scientific disciplines from the perspective of Objective Cognitive Constructivism, see Esteves et al. (2019) and Stern (2014, 2017b).     
  
Notwithstanding the complex nature of the historical development of either technologies or scientific theories, we argue that an important conclusion holds: their evolution can be characterized as an essentially  recursive development process converging, at least locally, to progressively perfected (eigen-) solutions. 

\subsubsection*{Objects are Tokens for Eigen-Solutions: \\ Precision, Stability, Separability and Composablility}

This subsection examines the nature and representation of recursive processes, the concept of eigen-solution an its essential properties,  and their importance for the justification of qualified forms of circular argumentation.

\begin{figure}[tb] 
	%\mbox{} \vspace{0mm} \mbox{}    
	
\centerline{ 
\includegraphics[height=48mm, %width=\linewidth, 
 trim=0mm 0mm 0mm 0mm, clip]{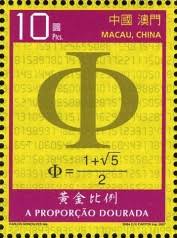}   
		%%trim=l b r t    	
  \mbox{} \hspace{1mm} \mbox{}  
\includegraphics[height=48mm, %width=\linewidth, 
 trim= 0mm 0mm 0mm 0mm, clip]{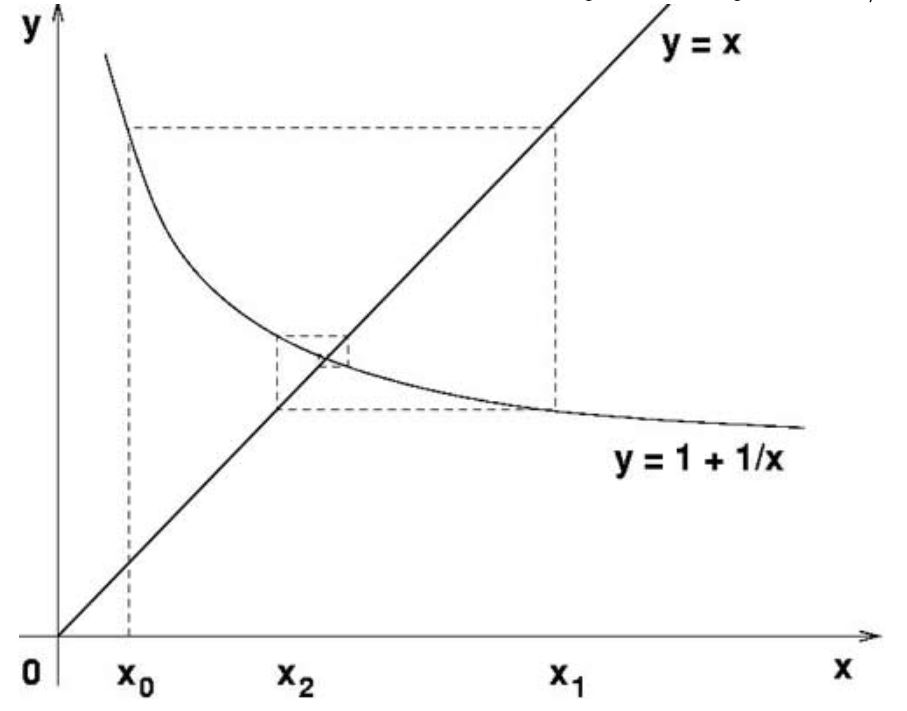}   
  \mbox{} \hspace{0mm} \mbox{}  %Phi07.png Phi02.jpg  
\includegraphics[height=48mm, %width=\linewidth, 
 trim= 0mm 0mm 0mm 0mm, clip]{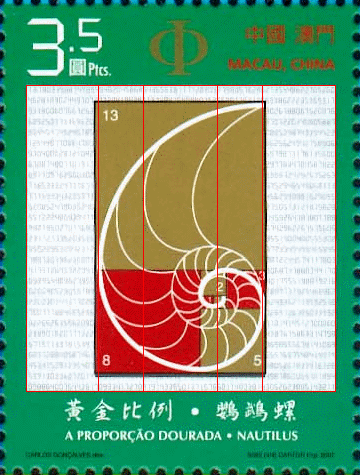}   
	} 
	
\caption{Fixed-point method converging to $\phi= 1 +1/\phi =  (1+\sqrt{5})/2$. }  

\end{figure}

Recursive computational methods are routinely used in applied mathematics.  
Figure 11 illustrates the \textit{Fixed Point iteration method}, based on the repeated iteration $x_{n+1}=f(x_n)$,  being used to compute $f(x)=1+1/x$, 
and converging to the \textit{golden ratio}, $\phi$, an irrational number, see Stern (2011b).     
Convergence to an invariant point, $x=f(x)$, is not guaranteed, only occurring under certain regularity conditions, see Garcia et al. (2003), Kelley (1987a,b) and Wimp (1984).  
In one dimension, such an invariant point can also be called an \textit{eigen-value}, and this concept can be generalized to multi-dimensional spaces, for example:  
An \textit{eigen-vector}, $x$, of a vector space operator $A$ (a matrix operator in ordinary linear algebra) is defined by the recursive relation 
$x= (1/\alpha) Ax$. 
Under appropriate regularity conditions, the set of such eigen-vectors constitutes a \textit{basis}, that is, any vector in the space can be decomposed into or recomposed from these basic eigen-vectors, see Sadun (2007). 
Well behaved invariants in  functional spaces are called \textit{eigen-functions} and, in general settings, such entities are called \textit{eigen-solutions}.   
This is the departing point for Heinz von Foerster (1911-2002) celebrated aphorisms:  

\begin{quotation}  
$\bullet$ \textit{Objects are tokens for eigen-behaviors}; \ and 

$\bullet$ \textit{Eigen-solutions are precise (sharp), stable, separable and composable}.     

\end{quotation} 
  
von Foerster used variant forms of this celebrated aphorism, see Foerster and Segal (2001) and Foerster (2003), to convey the following  ideas: \\
$\bullet$ The ``objects'' we perceive in our environment stand for invariant forms or fixed-points of our behaviors or ways of interaction with the same environment;  \\      
$\bullet$ Well-behaved (or objective) invariants are characterized by the four essential properties stated above, a  
topic to be further developed in Section 6. 

Circular argumentation is often identified as an indicator of fallacies, for it can be used to support all sorts of bogus statements. For example, a lier and his accomplice can argue: ``I never lie, as my friend can attest; and my friend opinion is reliable, for I vouch for him.''   
Under this circumstances, being aware of its intrinsically circular and recursive nature, what is the justification for using the Objective Cognitive Constructivism epistemological framework in the context of science? 

We claim that such a justification can be found in the four essential properties of eigen-solutions, namely: Precision (or sharpness), stability, separation and composition. 
Precision (or sharpness) is the hallmark of exact sciences, in fact, this is why they are called ``exact''. 
The great heroes of a scientific discipline can be easily found in a textbook's subject index: Their names are associated to the discipline's precise \textit{laws} or \textit{equations} (like Kepler's laws in astronomy, Newton's law in classical mechanics, Maxwell's law in electro-magnetism, Einstein's equations in special or general relativity, etc.).       
There is good reason for the status of these heroes and, more importantly, the high status of the corresponding eponymic laws or equations. 
More refined explanations concerning the four essential properties of eigen-solutions are given in the following sections. For now, it suffices to acknowledge that the precision of the aforementioned equations, together with their compositionality properties (that is, their analytic and synthetic power) are ultimately responsible for the success of exact sciences that, in turn, constitute the foundation of all subsequent technological applications that, in turn, allow the production of so many gadgets and wonders we enjoy in the modern world.  
    
This conjunction of, on one hand, the very high standards for the testing and validation of scientific laws imposed the four essential properties of eigen-solutions (precision, stability, separability and compositionality) and, on the other hand,  the extraordinary power of scientific laws is, we claim, the primary source of justification for accepting the circular characteristics of the Objective Cognitive Constructivism epistemological framework. 
In a nutshell, we argue that this combination of pragmatic power and high standard of validation provides a firm grip on reality, supplying reliable anchors for a scientific discipline;   
in this specific sense, the same scientific discipline can be considered \textit{objective}.  
In the context of mathematical statistics, the degree of precision and accuracy of a statistical hypothesis, together with the exactness and analytic/ synthetic power  of (de)composition rules associated to the corresponding eigen-solutions, are the key to define quantitative measures of \textit{objectivity}, as further discussed in Section 7.

Formally, a set of circular (coss-referent and recursive)  definitions can be handled rigorously using non-well-founded set theories, see Aczel (1988), Barwise and Etchemendy (1987), and Barwise and Moss (1996).    
Computationally, circularly defined data sets can be  represented, stored and manipulated using appropriate data structures, see Akman and Pakkan (1996), Iordanov (2010), and Pakkan and Akman (1995).     

Sections 6, 7 and 8 examine in further detail the role and characterization of eigen-solutions in the Objective Cognitive Constructivism epistemological framework. 
Meanwhile, Section 5 examines in further detail the role of causal explanations in the practice of science, with particular attention to some aspects of circular forms of argumentation that complement the discussion of this section.

\section{Know-Why and Effective Metaphysics}

In 1637 Descartes explained his refraction law, $\sin \theta_1 / \sin \theta_2$ $= n_2 / n_1$, using an analogy  based on the trajectory of bouncing tennis balls, see Figure 12 (left and center). 
In the case of reflexion, the horizontal (or tangential to the media's boundary interface) component of the ball's velocity remains unchanged, while its vertical (or normal to the interface) component is inverted. 
By analogy, in the case of (light's) refraction, Descartes assumed the tangential velocity to remain unchanged, while he assumed the total velocity to be a characteristic of the medium. 
Moreover, Descartes assumed the (total)  velocity of light to be \textit{proportional} to the medium's refraction index, see  McDonough (2015, p.552) and Mihas (2008). 
Hence, in Descartes' interpretation,   
$n_2/n_1=v_2/v_1$, 
implying that light travels faster in more refractive media. 
  
In 1662 Pierre de Fermat derived the same refraction law from the \textit{least time} principle, stating that a light ray traveling from point $P$ to point $Q$ takes the path that minimizes its travel time, see Figure 12 (right).  Unlike Descartes, Fermat assumed that the speed of light is \textit{inversely proportional} to the medium's refraction index. 
Hence, in Fermat's interpretation,    
$n_2 /n_1=v_1/v_2$, 
implying that light travels slower in more refractive media.  
Appendix B gives a simple derivation of Descartes' refraction law from Fermat least time principle using the language of modern calculus.

\begin{figure}[bt] 
	%\mbox{} \vspace{0mm} \mbox{}       
	
	\centerline{ 
		\includegraphics[height=1.7in]{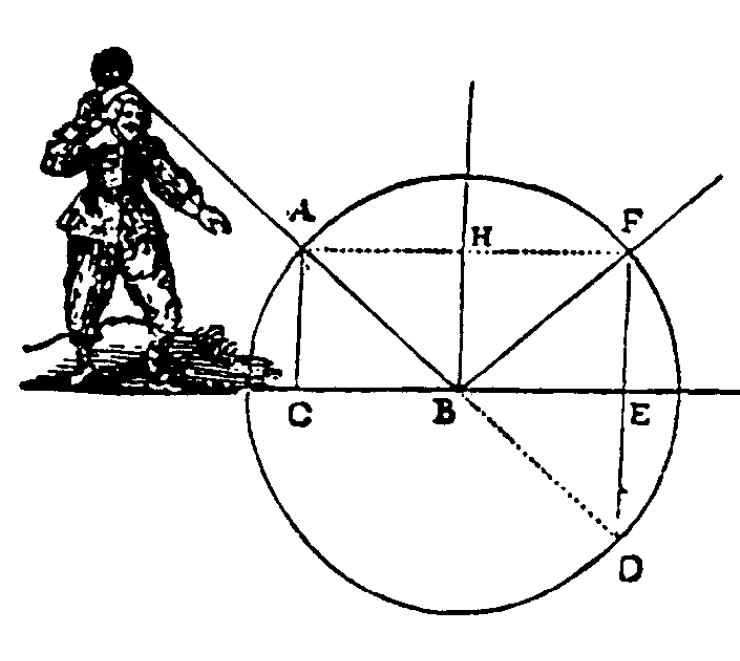}
		\mbox{} \hspace{-1mm} \mbox{} 
		\includegraphics[height=1.7in]{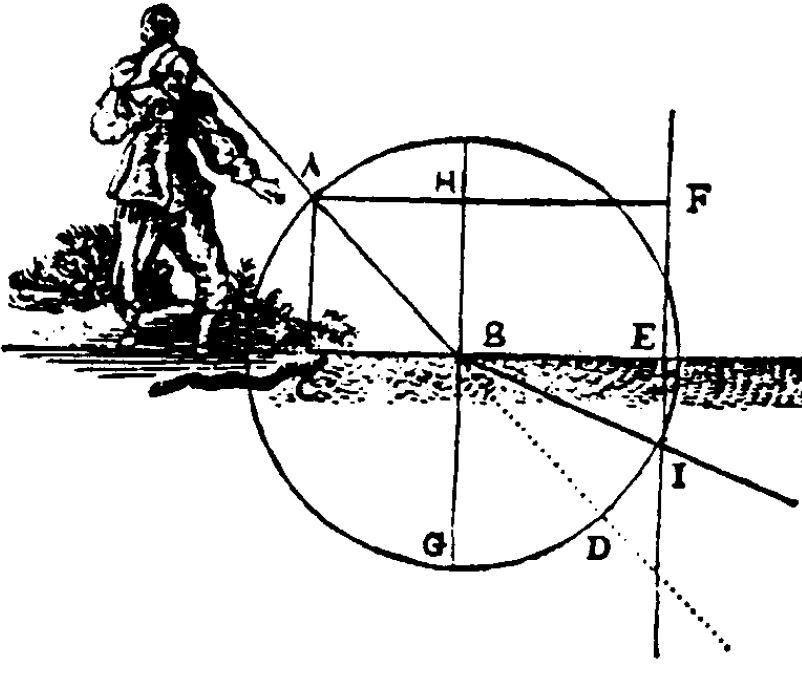}
		%\mbox{} \hspace{1mm} \mbox{} 
		%\includegraphics[height=1.5in]{Optics73.jpg} 
		\mbox{} \hspace{1mm} \mbox{} 
		\includegraphics[height=1.75in, width=1.2in]{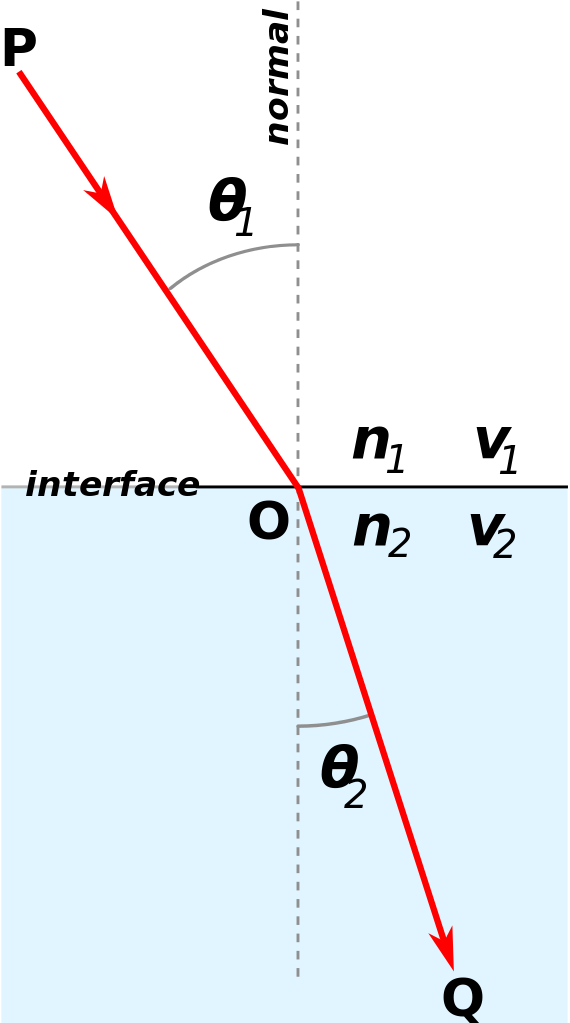}  
	}%centerline 
	
	\caption{Tennis ball analogy for Descartes' (1637) law}  
	
\end{figure}

Notice that both Descartes and Fermat worked long before any measure of the speed of light was available.  
The first quantitative estimate of the speed of light, in sidereal space, was obtained in 1976 by Ole Roemer (1644-1710), 
making use of what we would today call a ``Doppler effect'' on the apparent period of Io (a satellite of Jupiter discovered by Galileo in 1610), 
see Shea (1998) for further details and a historical contextualization of Roemer's work. 
%More precisely, he measured the violet and red shifts  (i.e. variations for shorter and longer apparent periods) of Io, as observed from Earth traveling in its orbit towards and away from Jupiter. 
%Roemer's final estimate was $c=1au/11'$, that is, one astronomial unit (the length of the semi-major axis of the earth's elliptical orbit around the sun, approximately 150 million kilometers) per $11$ minutes;  today's value is around $1au/8'20''$. 
Almost two centuries latter, in 1850, the first direct measurements of the comparative speed of light in distinct material media (like air and water) were obtained  by L\'{e}on Foucault (1819-1868) using a rotating mirror device, see Tobin (1993) and Jaffe (1960) for further details. 
Only these latter measurements could finally settle the old controversy concerning the faster or slower speed of light in more refractive media, favoring Fermat over Descartes.  

The aforementioned examples by Descartes and Fermat  are fine examples of \textit{metaphysical} models, where the adjective metaphysical is used in two distinct but complementary senses: 
First, in the \textit{literal} sense, a meta-(beyond)-physical model makes reference to non-observable entities or non-observed quantities like, in the case at hand, the speed of light.       
Second, in the \textit{gnoseological} or Aristotelian sense, a metaphysical model provides answers for \textit{why} questions, that is, it provides explanations for why things are or behave the way they do.  

Notwithstanding the practical impossibility of observing the speed of light at the time of Fermat and Descartes, their models offered contradictory predictions concerning the proportionality or inverse proportionality relation between the speed of light and the propagation medium's refraction index.  
Moreover, these contradicting predictions awoke the curiosity of scientists, sparkled the imagination of philosophers and drove centuries of research in the field of optics, including the development of technological means and methods for measuring the speed of light, as well as the development of alternative models and theories concerning the nature and properties of light.     
For historical perspectives of several competing theories of light we refer to Ronchi (1970), Sabra (1981) and Zempl\'{e}n (2005). 

Descartes' model focus on \textit{efficient} or moving ($\kappa \iota \nu o \upsilon \nu$) causes, that is, it focus on the agents that interact in a process to make something behave the way it does.  
In the model at hand, distinct propagation media interact with tennis balls or light rays so that they are forced to change their velocity in a certain way (given specific initial conditions).  
In contrast, Fermat's principle offers a \textit{teleological} or \textit{final}  cause, that is, it explains a process by the goal or end ($\tau \epsilon \lambda o \varsigma$) at which the process is directed.    
In the model at hand, the goal is to minimize the travel time of the light ray (given specific boundary conditions). 
These \textit{causes} or \textit{explanations} ($\alpha \iota \tau \iota \alpha$) are very different in nature, for example: Efficient causes refer to events occurring before what is being explained, that is, they are based on prior conditions to the explanandum. In constrast, teleological explanations refer to events occurring after what is being explained, that is,  they are based on posterior conditions to the explanandum.

Fermat's principle is expressed minimizing total travel time; mathematically, this principle can be formulated by minimizing an integral of the (differential) quantity $ds=1\; dt$.
In a similar way, Leibniz, Euler, Mauperius, Lagrange, Jacobi, Hamilton, and many others were able to reformulate Newtonian mechanics, minimizing the integral of (differential) {\it action},  
$ds= L(t)\; dt$, where the Lagrangian, $L(t)$, is the difference between the kinetic energy (Leibniz' vis viva), $(1/2)mv^2$, and the potential energy of the system (Leibniz' vis morta). 
Hence, these formulations are called in physics principles of minimum action or \textit{principles of least action}; for an elementary text see Lemons (1997), for standard  introductions see Goldstein (1980), Krasnov et al. (1973) and Marion (1970), for historical accounts see Dugas (1988) and Goldstine (1980), for philosophical discussions see Nagel (1979), Tomasello (2005), Yourgrau and Mandelstam (1960) and Woodfeld(1976).   

At the end of the XIX century, least action and closely related teleological principles became the grand unifying theme  for theoretical physics -- in the words of Helmholtz (1887),  a \textit{leitmotif} dominating the whole of physics -- and the accompanying methods of variational calculus became one of the most powerful tools of mathematical physics and other natural sciences, see Yourgrau and Mandelstam (1960, ch.14). 
Notwithstanding the enormous importance of least action principles in modern science, there are persistent philosophical suspicions or even outright distrust concerning the use of teleological principles.               
The reason for these suspicions are related to the possibility of assuming causal influences traveling backwards in time, leading to fallacies closely related to the vicious forms of circular argumentation discussed in the previous section, see Nagel (1979),  Terekhovich (2012) and Woodfleld (1976). 

One way out of the possible circularity implied by  teleological principles is to invoke a prior existing mind, consciousness, controller or similar entity that is able to foresee or take into consideration (either perfectly or with some uncertainty) all possible evolution paths that could be taken by the system, and then choose (by setting appropriate decision variables) the path that will  best attain or come closer to the stipulated goal.    
However, in many applications, like the laws of physics, such an assumption may seem strange or spurious, see Helrich (2007). 
Even in the case of a biological organisms, where such assumptions may be natural, we may not know in sufficient detail the inner workings of the postulated controllers, making them weak providers of efficient causes. 

The literature of Cybernetics and System's Theory offers some epistemological settings conceived to overcome the aforementioned difficulties related to the use of teleological principles, some of which could be seen as precursors to ideas discussed in the previous section. 
We proceed with a brief exposition and comparative analysis of a specific example, in hope of further clarifying the perspective offered by the Objective Cognitive Constructivism epistemological framework.

Gerd Sommerhoff (1915-2002) proposes the notion of teleology as an abstract but objective systemic property that can be discovered and attested by observing the behavior of a system (over time), without postulating any conscious purposiveness or hidden mental states.  Sommerhoff is primarily interested in living organisms and cybernetic systems, but his concepts can easily be adapted to physical systems; in this case, his terminology of variables in the system's \textit{environment} and \textit{action} spaces could be replaced by even more neutral terms like input and output (or answer) variables.     
In a nutshell, Sommerhoff (1969; 1990, ch.3) notion of \textit{goal-directedness as an objective system-property} can be summarized as follows: 

(1) \textit{If an action is directed towards a goal $G$, then [...] there exist at at least one point in time $t_k$ during the action, two variables $e$ and $a$ defined on the environment and the action [space] respectively, and a function $f$ such that $f(a_k,e_k)=0$ is a necessary condition for the subsequent occurrence of $G$.} Sommerhoff (1969, p.167). 

(2) \textit{The variables $a$ and $e$ in the funcion $f$ introduced above are mutually orthogonal [...that is...]  $f(a_k,e_k)=0$ is a condition that must be \emph{brought about} by the mechanism involved. It is not implied by the axioms of the system. Indeed, random combinations of $a$ and $e$ are conceivable initial states of the system.} Sommerhoff (1969, p.165). 

(3) \textit{The results produced by these directive processes have a high degree of a priori improbability. Indeed, it can be said that the sense of wonder we feel ... is essentially a surprise reaction evoked by the seeming improbability, and hence unexpectedness, of the phenomena we witness. And in each case these phenomena have come about, despite their inherent improbability because they have been the end-product of a directive activity of one kind or another...} Sommerhoff (1990, p.50).

Let us consider the three conditions above stated by Sommerhoff. If $a$ and $e$ are variables in a continuous space (like real numbers or vectors), the  \textit{equation} $f(a,e)=0$ defines a \textit{precise} subset of the same space (technically, a lower-dimensional algebraic sub-manifold). 
Such a subset is \textit{sharp}, in the sense that it is very hard to be reached by pure chance (technically, it is a subset of zero Lebesgue or natural probability measure). 
Nevertheless, the laws of physics (like the refraction law) are stated in this exact manner. 
Furthermore, we can attest how the phenomena described by these laws ``come about despite their inherent improbability''. 
In the Objective Cognitive Constructivism epistemological framework this situation is known as the \textit{Zero Probability Paradox} --  regarding the possibility of attesting strong statistical support for a statement describing an extremely improbable (zero measure) event. 
For an extended discussion of the Zero Probability Paradox and its epistemological consequences, see Stern (2011b); for the technical difficulties it poses to traditional methods for test of hypotheses in statistical science and how these difficulties are overcome by the $e$-value significance measure, see Stern and Pereira (2014) and Esteves et al. (2019). 

Before ending this section, let us make a few comments on the importance of metaphysical models, focusing on the use of alternative models even when they do not directly imply discrepant empirical consequences.    
Richard Feynman (1964, ch.19) analyses three different but mathematically (almost) equivalent ways of formulating the classic law of gravitation, namely, using Newton's attraction forces, the local field method, and the principle of least action. Feynman (1985, pp. 50-53) says:    

\textit{ 
	%I am talking about the theories [that] are exactly equivalent. 
	Mathematically each of the different formulations [...]  gives exactly the same consequences. 
	%What do we do then? You will read in all the books that we cannot decide scientifically on one or the other. 
	That is true. They are equivalent scientifically 
	%It is impossible to make a decision, because 
	[...] there is no experimental way to distinguish between them if all the  consequences are the same. 
	But psychologically they are very different in two ways. 
	First, philosophically you like them or do not like them; and training is the only way to beat that disease. 
	Second, psychologically they are different because they are completely unequivalent when you are trying to guess new laws.   
	As long as physics is incomplete, and we are trying to understand the other laws, then the different possible formulations may give clues about what might happen in other circumstances. 
	In that case they are no longer equivalent, psychologically, in suggesting to us guesses about what the laws may look like in a wider situation.} 

Richard Feynman offers the view of a great scientist pushing the state of the art in physics, a task he was able to successfully accomplish several times; for a prime example of his accomplishments related to formulation of Quantum Mechanics by teleological principles, see Feynman and Hibbs (1965). I will add some remarks on the importance of considering alternative metaphysical models and mathematical formulations of physical laws from the humble perspective of a physics teacher:

Occasionally, I had the responsibility of teaching 
an introductory discipline in Analytical Mechanics to 
junior students of physics or applied mathematics, using standard textbooks like Marion (1970) or Goldstein (1980). 
One of the first tasks in such a discipline is to 
prove that (except for some exceptional cases at the  
corners of the formalism) the laws of mechanics, 
as described by Newton, Lagrange or Hamilton, 
are exactly the same. 
However, after achieving this goal, 
I emphatically warn the students that such theorems 
can be quite misleading. 
More specifically, I try to show them that, if they 
want to harness all the power provided by the tools 
of Analytical Mechanics, they should be willing to 
embrace full-heartedly its teleological spirit, and   
get ready to ask and imagine --  what is it that the 
particle or the spin-top ``should do'' or ``wants to do''  --  in order to minimize the appropriate functional of its (future)  trajectory.  
This is not an easy skill to master, being perhaps even more challenging then the skills needed for correct algebraic manipulation of the pertinent equations. 
Nevertheless, from my experience, developing this 
sort of geometric view and analogical thinking is 
the very key to success in this discipline, for it provides the guidelines to approach each new problem,  
to correctly structure its mathematical framework, and to assemble the equations required to solve it.

\section{Know-How and Well-Adapted Ontologies}    

As a teenager, in the early 70's, I had the opportunity to take classes at the S\~{a}o Paulo municipal planetarium, and also to participate at the local Amateur Telescope Making club. 
Figures 13 (left) illustrate the activity of grinding your own telescope mirror, a task accomplished by rubbing two initially plane glass discs against each other  with the help of a sequence of coarser (for grinding) to finer  (for polishing) abrasive powders.  
In this process, the disc at the top, the tool, takes a convex shape, while the disc at the bottom, the mirror, takes a concave shape.   
The goal of this long and tedious work is to obtain a perfectly spherical (or parabolic) and well polished mirror piece, that can then receive a light reflective silver  coating, in order to be used as a mirror for an actual amateur telescope, see Ingalls (1953), Thompson (1947) and Texereau (1957).

\begin{figure}[bt] 
	%\mbox{} \vspace{0mm} \mbox{}     

\centerline{ %trim=l b r t  
	\includegraphics[height=1.5in, width=1.15in, 
	trim= 0mm 0mm 0mm 0mm, clip]{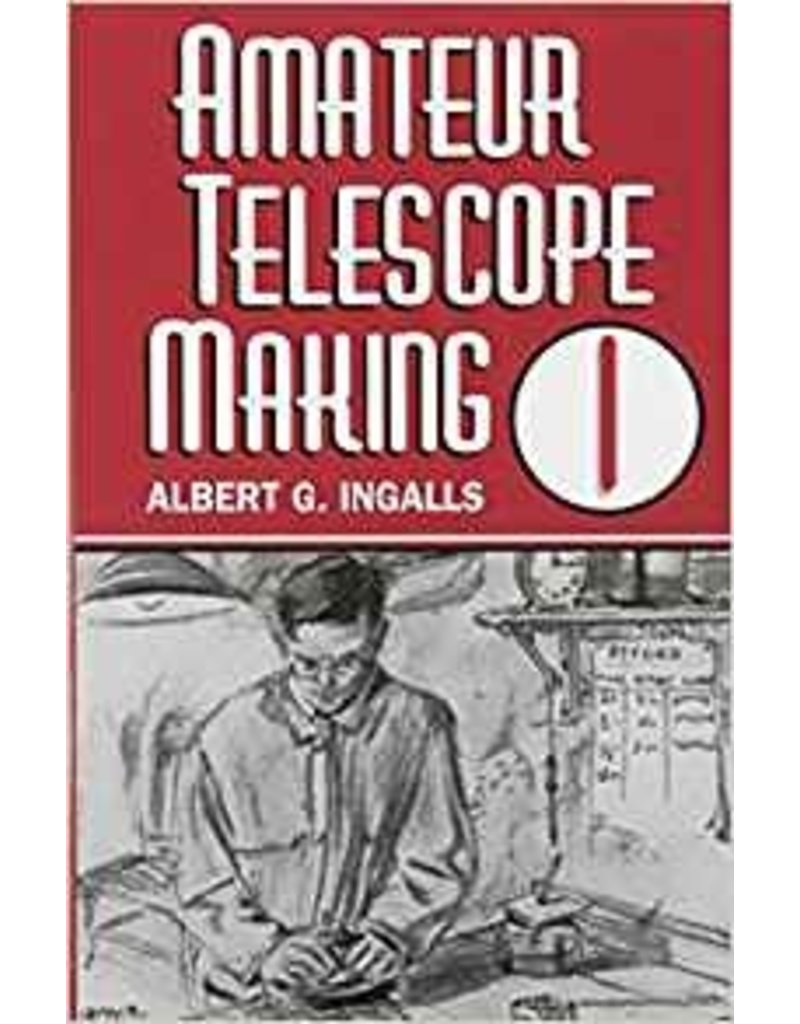}  
	\mbox{} \ 
	\includegraphics[height=1.5in, width=1.1in, 
	trim= 0mm 0mm 0mm 0mm, clip]{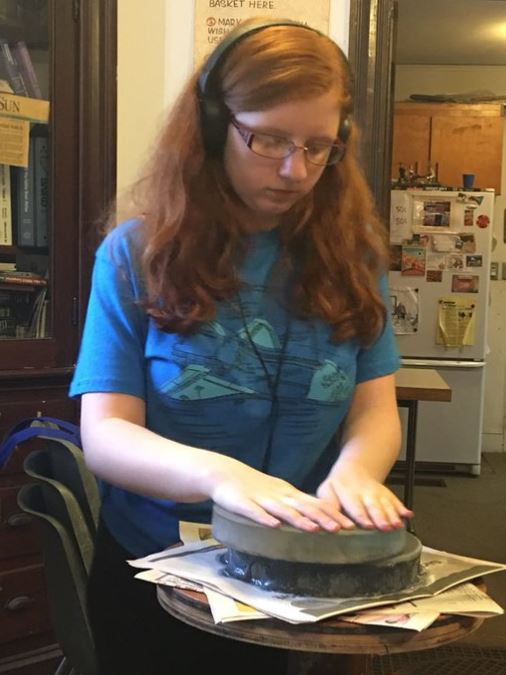}
	\mbox{} \ 
	\includegraphics[height=1.5in, width=1.05in, 
	trim= 0mm 0mm 0mm 0mm, clip]{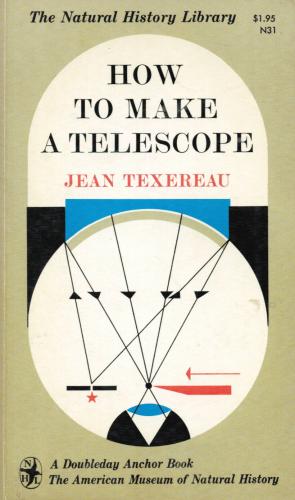} 
	\mbox{} \ 
	\includegraphics[height=1.5in, width=1.05in, 
	trim= 0mm 2mm 0mm 0mm, clip]{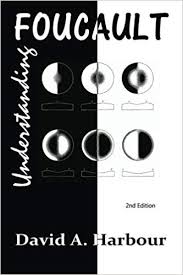} %{HarpTx8.jpg}
} %trim=l b r t  
  	
\caption{Grinding, testing  \& adjusting a telescope mirror precise shape}  
   
\end{figure}

All points on the surface of a perfectly spherical lens or mirror should be at the same distance of its center of curvature. 
However, actual grinding is not a flawless process, and real lenses or mirrors deviate from their ideal forms. 
Consequently, the surface of an actual mirror or lens wabbles around its ideal form, like the waves exhibited by a thin disc plate vibrating around its natural resting plane, or the ripples we see at the surface of the cup of coffee in Figure 14 (center). 
%%%% 
Figures 14 (left and right) show typical angular and radial profiles of these characteristic modes of vibration.

\begin{figure}[b] 
	%\mbox{} \vspace{0mm} \mbox{}     

\centerline{  
	%\includegraphics[height=1.2in,  %trim=l b r t 
	%trim= 0mm 0mm 0mm 0mm, clip]{HarpD4.jpg}  
	%\mbox{}  
	\includegraphics[height=1.6in,    
	trim= 2mm 2mm 2mm 2mm, clip]{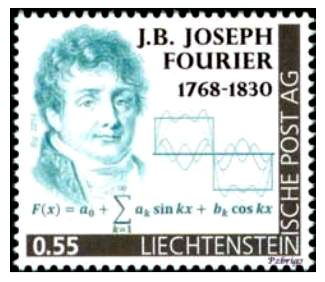}  
	\mbox{} \ 
	\includegraphics[height=1.6in, width=1.65in, 
	trim= 21mm 4mm 20mm 4mm, clip]{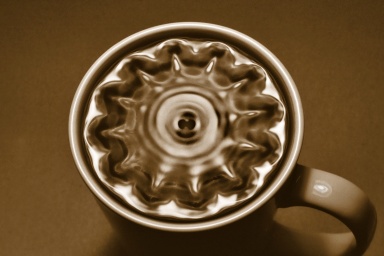}  
	\mbox{} \   
	\includegraphics[height=1.6in,    
	trim= 20mm 5mm 24mm 5mm, clip]{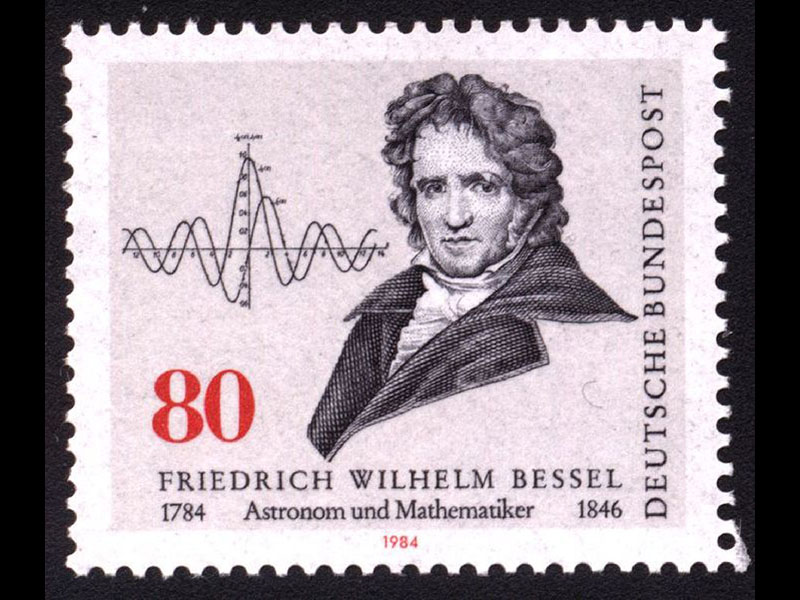} 
    %\includegraphics[height=1.2in]{Lamp28.jpg} 
    %\mbox{} %trim=l b r t  
}%centerline 
  	
 \caption{Waves and wobbles on the surface of a disc}  
  	
\end{figure}

The aforementioned characteristic forms of vibration can be mathematically described by special eigen-functions.  
Figure 15, a composite from Genberg and Michels (2004) and Zou and Wattellier (2012), gives, at the top row, an illustration of a few of these eigen-functions, that also describe typical forms of \textit{aberration} (i.e., deformation relative to its ideal form) that may occur in the surface of a mirror.   
The next subsection comments on some relevant  mathematical properties of these eigen-functions.

\begin{figure}[bt] 
	%\mbox{} \vspace{0mm} \mbox{}     
	
\centerline{ %trim=l b r t  
	\includegraphics[width=1.5in, height=1.1in, 
	trim= 0.5mm 0mm 0mm 0mm , clip]{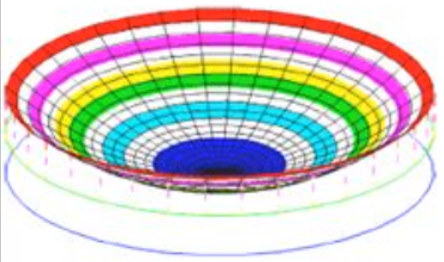} 
	\mbox{}  
	\includegraphics[width=1.5in,  height=1.1in]{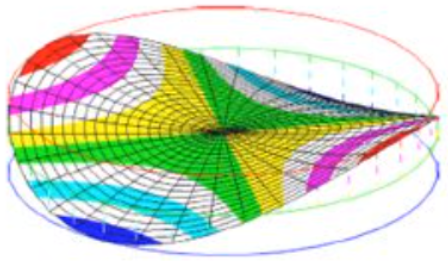} 
	\mbox{}  
	\includegraphics[width=1.5in,  height=1.1in]{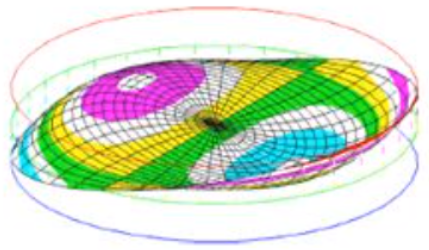} 
	\mbox{}  
	\includegraphics[width=1.5in,  height=1.1in]{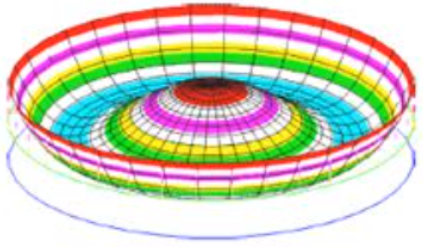} 
}%centerline 

\mbox{} \vspace{0mm} \mbox{} 

\centerline{ 
	\includegraphics[width=1.5in, height=3.1in]{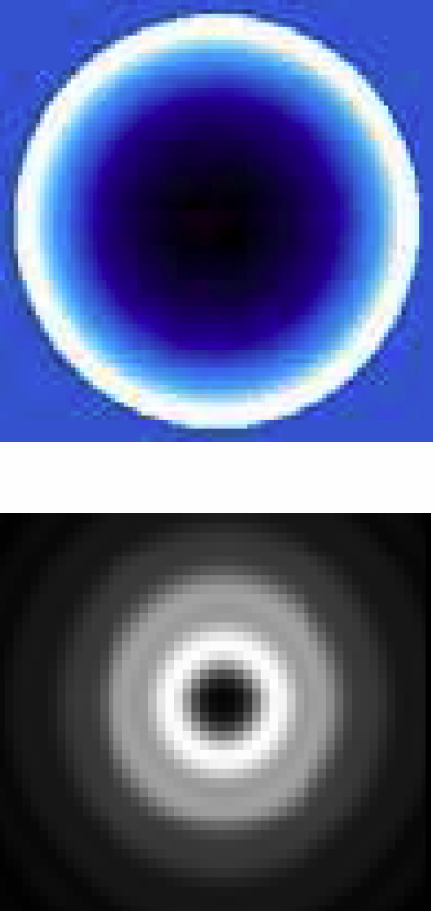} 
	\mbox{}  
	\includegraphics[width=1.5in, height=3.1in]{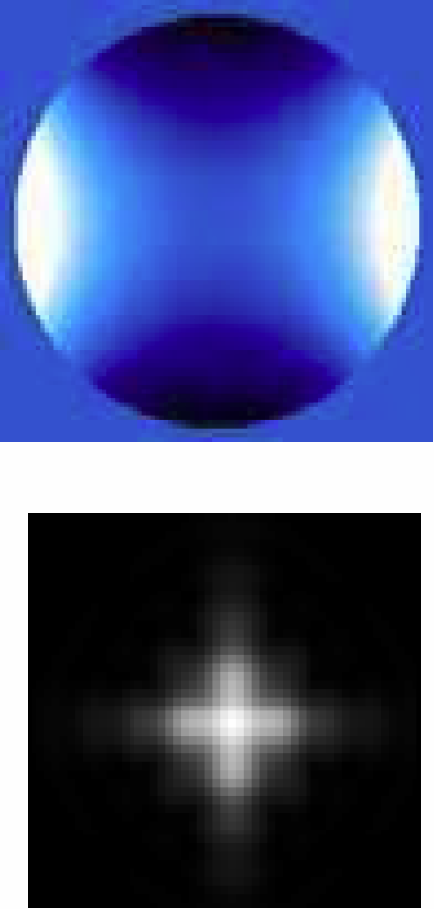} 
	\mbox{}  
	\includegraphics[width=1.5in, height=3.1in]{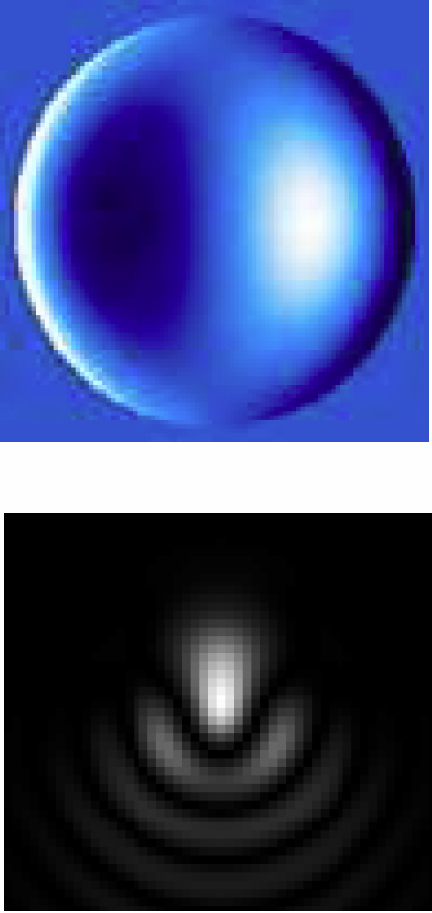} 
	\mbox{}  
	\includegraphics[width=1.5in, height=3.1in]{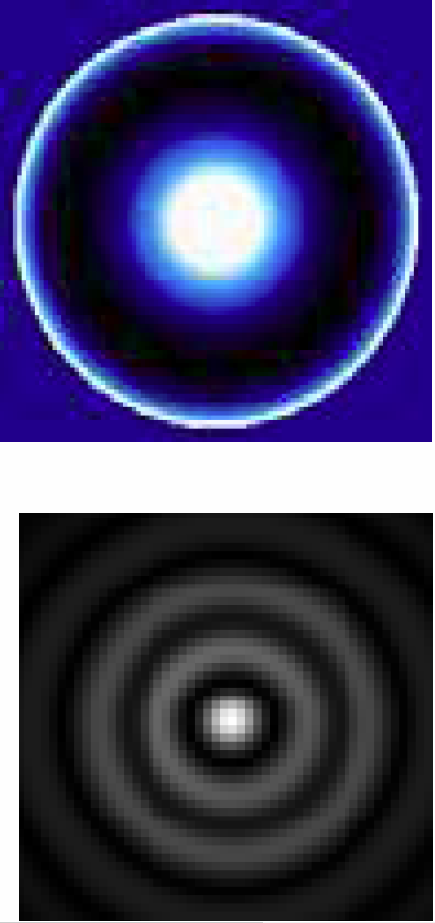} 
}%centerline 

%\mbox{} \vspace{-0mm} \mbox{} 

%\centerline{Figure 14: $Z(n,m)$ - Zernike eigen-functions/  objects in optics/ ophthalmology:}  
%\centerline{$Z(2,0)$, Defocus; $Z(2,2)$, Astigmatism; $Z(3,1)$, Coma; $Z(4,0)$, Spherical.}  

\caption[]{$Z(n,m)$  -- Zernike eigen-functions and  objects in ophthalmology and  \\%\hspace{\textwidth}   
 optics: $Z(2,0)$, Defocus; $Z(2,2)$, Astigmatism; 
	$Z(3,1)$, Coma; $Z(4,0)$, Spherical.}  
    
\end{figure}

Thank God, as a teenager, I did not need to know all that heavy mathematics in order to make my amateur telescope. 
Figures 13 (right) and Figure 15 (middle row) illustrate a practical technique (developed by the already mentioned L\'{e}on Foucault) used to carefully illuminate the surface of the mirror piece during the grinding process in order to obtain light and shadow effects that allow the detection of the aforementioned aberrations, see Harbour (2001) and Texereau (1957). 
Once a characteristic aberration is detected, a corresponding corrective grinding maneuver (a.k.a. retouching procedure) can be applied, allowing the mirror fabrication process to converge to the desired ideal form, see next sub-section for additional details.

Historically, see Abrahams (1994) and King (1955), practical know-how used for telescope making was developed by optical artisans long before the development of the corresponding mathematical eigen-function theory, even though, as it can clearly  be seen in Figure 15, there is a close connection between the two.

Figure 15 (bottom row) depicts the image of a single 
and distant bright point produced by an optical element (mirror or lens) suffering from the corresponding aberration. 
The reader will probably recognize the standard names 
given to these eigen-forms as ``diagnostics'' made by 
an ophthalmologist or optometrist, for these names correspond to typical aberrations affecting the biological lens of an human eye.  
Long before the invention of the telescope, since the late XIII century, artisans manufacturing magnifying glasses and spectacles had a practical know-how about these eigen-forms, being able to manufacture  eyeglasses with the same or complementary characteristics that, in turn, were used to alleviate corresponding symptoms of their patients. 
In fact, it can be argued that it was the ready availability of a variety of such lenses that made the invention of the telescope possible in the XVII century, see Helden (1977), Helden et al. (2010), Ilardi (2007) and also Morrison (1987), including the episode \textit{Lenses and Telescopes} from the accompanying PBS videos.       

Now the stage is set for discussing ontology, from the perspective of the Objective Cognitive Constructivism epistemological framework. 
 
$\bullet$ In computer and information science, an \textit{ontology} provides a basis for knowledge representation and information processing in an  established field of activity, like a scientific discipline.
It can be conceived as a formal lexicon, that is, a controlled vocabulary with nouns used to name  key entities and concepts of the pertinent domain of discourse, and also adjectives, verbs and adverbs used to name their properties and interrelations.     
   
$\bullet$ In classical philosophy, the word ontology is etymologically derived from $o\nu$ or $\epsilon o \nu$ (being, that which is, actual, real), the present participle of the verb $\epsilon \iota \mu \iota$  (to be). Hence, ontology concerns the study of being, that is, it addresses the question of what entities exist or may be said to exist, their properties and interrelations. 
  
In the Objective Cognitive Constructivism epistemological framework, both notions of ontology are brought together by the key notion of \textit{objectivity}, as discussed in previous sections.  
Recalling once again von Foerster's  metaphor:  
\textit{Objects are tokens for eigen-behaviors}. 
Furthermore, eigen-solutions have four essential properties, namely, precision (or sharpness), stability, separation and composition. 
Finally, the quality (or ontological status) of  objects can be evaluated and measured by the degree in which the corresponding eigen-solutions manifest their four essential properties, as further discussed in the next subsection.

\subsubsection*{Orthogonality, Completeness, Invariance and  Well-Adaptedness} 

This subsection discusses the four essential properties of eigen-solutions as they are expressed, in a very strong and rigorous way, by the eigen-functions discussed in this section.    
Some details in this subsection presuppose familiarity with pertinent mathematical methods; however, these details can be skipped without impairing the understanding of general ideas.  

The characteristic (angular and radial) eigen-forms of a vibrating disc plate are mathematically described by eigen-functions traditionally named after, respectively, Joseph Fourier (1768-1830) and Friedrich Bessel (1784-1846), see Butkov (1968),  Courant and Hilbert (1953),  and Figure 14. 
 %%%  
The corresponding joint two-dimensional  description of the plate's vibration is known as a Fourier-Bessel decomposition, see  Butkov (1968, Sec.9.7, p.363-368),  Courant and Hilbert (1953, v.1, Sec.5.6, p.307-308), Reddy (2007, Sec.5.5.1-2, p.200-202), and Figure 15.  
 %%% 
In the field of optics, Zernike polynomials, $Z(n,m)$, are commonly used as a computationally more convenient representation but mathematically equivalent alternative to the  Fourier-Bessel decomposition, see  Khotanzad and Hong (1990), Lakshminarayanana and Flecka (2011) and Wang et al. (2008).

\textit{Orthogonality} is a strong form of separability, and a most convenient property  exhibited by the eigen-fuctions used in the case at hand.  
 %%% 
Orthogonality means that we can apply a retouching maneuver designed to correct a single characteristic aberration, without aggravating another, that is, it is possible to deal  \textit{separately} or independently with simple problems corresponding to each element  of the eigen-function basis.  
 %%%  
From a physical point of view, 
each element of the Fourier-Bessel decomposition describes a standing, self-sustaining or resonating mode of vibration. Accordingly, each of these eigen-forms of vibration can store a well-defined amount of energy, see French (1971, Ch.6, p.181-188) and Stern (2017a, Sec.5).

The concept of orthogonality can be relaxed and usefully extended into many application areas, for example:  In the design of formal languages, it may designate non-redundancy  conditions; in categorization or classification systems, it may indicate mutually exclusive or non-ambiguity conditions; in combinatorics, it may reveal a matroid structure; in dynamic systems, it may imply local transversality  conditions; in statistical modeling, it may refer to non-correlation or alternative forms of non-association; etc.   

As in other examples previously examined in this article, either the Fourier-Bessel or the Zernike basis of eigen-functions is \textit{complete}, that is, any (continuous) deformation of a disc plate can be composed as a linear combination of its elements.  
 %%%
In this way, one can decompose a complex aberration into its basic elements, allowing for separate corrective actions that, together, take care of the original complex problem.

Well-Adaptedness (to the specific context of application) is yet another important property of the eigen-functions under consideration. 
The availability of an orthogonal system of eigen-forms of aberration and corresponding corrective maneuvers would be of little help if we had to apply an infinite number of them. 
In principle, this unfortunate situation could be possible, 
for there are an infinite  number of those eigen-forms, corresponding to the eigen-functions 
$Z(n,m)$ for $n,m= 0,1,2,3\ldots$ . 

Fortunately, in the art of mirror grinding, only a few of those infinite eigen-forms of aberration are of practical importance, namely, only eigen-forms or \textit{lower order} (that is, terms corresponding to small values of $n$ and $m$) appear in practice.  
  %%% 
This kind of \textit{fast convergence} is a natural consequence of the specific eigen-functions being used and of energy conservation and finiteness, namely: The energy stored in angular  (or radial) normal vibration modes of order, $n$, that is, corresponding to  Fourier (or Bessel) terms or order, $n$, is proportional to the square of the vibration order, $n^2$, and the square of the vibration amplitude, $A_n^2$,  see Jarvis (2016) and Sadun(2007). 
Hence, the amplitude of higher modes of vibration must converge rapidly (super-linearly) to 
zero, technically: 
  
\[ E = \sum\nolimits_{n=0}^\infty E_n < \infty 
	\Rightarrow \sum\nolimits_{n=0}^\infty n^2 A_n^2 < \infty 
	\Rightarrow A_n = \mathcal{O}(\,1/ n^{3/2+\delta}\,) \, . 
\]

Fourier, Bessel and similar special functions provide systems of eigen-solutions that are well-adapted to many application areas of classical and modern physics, see Butkov (1968) and Courant and Hilbert (1953). 
Hence, following a path similar to the one 
taken in this article, it is possible  
to establish connections between 
elements of such a system of eigen-functions and key elements of the   scientific ontology  used in the corresponding application area. 
For example, Stern (2008, 2014, 2017a) analyze, from the perspective of Objective Cognitive Constructivism, the correspondence between key elements of musical ontology (like musical notes, chords and forms of harmony) and the eigen-solutions of a vibrating string, see also Benade (1960). 
Recent and ongoing research in wavelet theory aims to provide well-adapted eigen-solutions to a wider class of problems based on concepts like self-similarity and other recursive relations, see Aboufadel and Schlicker (1999) for a gentle introduction and also Resnikoff and Wells (1998).

\section{Heavy Anchors: Bayesian Epistemic Values}

The main objective of this article is to use intuitive ideas to provide an easy introduction to the objective cognitive constructivism epistemological framework. 
Accordingly, several aspects have been presented in (over) simplified form, for one can not hope to write a gentle introductory text striving, at the same time, for utmost rigor and completeness.  
In contrast, the main objective of this section is to consider some heavy methodological anchors developed in previous publications, for they are the tools used to build strong and well-defined connections between intuitive but somewhat intangible epistemological ideas and the practical procedures of empirical science.  
Additionally, this section surveys some previous related publications examining more technical or narrowly focused aspects of epistemology. 
   
Notwithstanding its introductory and intuitive character,  this paper is conceived as an integral part of a comprehensive research program in the area of Bayesian statistics and epistemology. 
This section presents a general overview of this research program, mapping several already published articles in which specific topics of interest related to the present paper are discussed in greater detail and depth.     
 
The Full Bayesian Significance Test (FBST) is a statistical test of hypothesis published by Pereira and Stern (1999) and further extended in Madruga et al. (2003), Pereira et al. (2008) and Esteves et al. (2016, 2019). 
This solution is anchored by a novel measure of \textit{statistical significance} known as the $e$-value, $\ev(H\g X)$, a.k.a. the \textit{evidence value} provided by observational data $X$ in support of the statistical hypothesis $H$ or, the other way around, the \textit{epistemic value} of hypothesis $H$ given the observational data $X$. 
The $e$-value, its theoretical properties and its applications have been a topic of research for many collaborators around the world for the last 20 years; Pereira and Stern (2020) give a panoramic view of this program, listing hundreds of publications.

\begin{figure}[tb]  
	\centering  
	\begin{small}  
		\begin{tabular}{c c c c c} 
			\cline{1-1} \cline{5-5}  
			Theoretical & \bvert & Metaphysical  & \bvert & Observational \\    
			\cline{3-3}  \\ 
			Statistical  & $\Rightarrow$ & Causal  
			&  $\Rightarrow$ &  Hypotheses  \\ 
			 models   & & explanation   & &  formulation    \\ 
			$\Uparrow$   & &               & & $\Downarrow$        \\  
			Speculative  &  \multicolumn{3}{c}{Sharp$^*$/ precise} & Experimental  \\ 
			interpretation &  \multicolumn{3}{c}{hypothesis validation} & (trial)  design  \\     
			$\Uparrow$   & &               & & $\Downarrow$         \\ 
			Statistical   & & Data & &  Experimental    \\ 
			analysis & $\Leftarrow$ &  processing     
			& $\Leftarrow$ &  observations \\  \\ 	
			\cline{3-3}     
			%\multicolumn{2}{l|}{Sample space} & Operational & 
			%\multicolumn{2}{|r}{Parameter space} \\ 
			Parameter space ($\Theta$) & \bvert & Operational & \bvert & 
			Sample space ($X$) \\ 
			\cline{1-1}  \cline{5-5}  \\ 
			\multicolumn{5}{c}{ {\bf Figure 16:} 
				Statistical modeling diagram.} 
		\end{tabular} 
	\end{small} 
\end{figure} 
\addtocounter{figure}{1}

Eigen-solutions, as depicted at the center of the Scientific production diagram in Figure 10, are represented, in the context of mathematical statistics, by sharp statistical hypotheses, as depicted at the center of the Statistical modeling diagram in Figure 16; see Stern (2017) for further details.   
The $e$-value and the FBST were specially designed to evaluate the degree of precision of sharp statistical hypotheses, and also to verify the accuracy of exact compositional rules for  hypotheses of this kind. 
Therefore, the $e$-value and the FBST can be used as a statistical touchstone that is taylor-made to the task of accessing the \textit{objectivity} of scientific eigen-solutions, as this notion is defined in the context of objective cognitive constructivism.  
Further mathematical, statistical, and logical or compositional properties of the $e$-value are discussed in Borges and Stern (2007), Esteves et al. (2016, 2019), Izbicki and Esteves (2015), and Stern et al. (2017).  

The Classical or Frequentist school of mathematical statistics offers the $p$-value as its signature significance measure for statistical hypotheses.  
In contrast, traditionally, the Bayesian school offers the Bayes Factor as its own alternative to a significance measure.   
These two were the leading schools of statistical science during the XX century, and each one of them has developed its own inference procedures for testing hypotheses. 

Nevertheless, the mathematical, statistical,  and logical properties of $e$-values, $p$-values and Bayes Factors are very different, and so are the properties of their corresponding hypothesis tests and related inference methods. 
Therefore,  the epistemological frameworks used by these three programs are also quite different, for they have to accommodate discrepant properties of the aforementioned significance measures. 
 % 
%% The present article, together with Stern (2017), aims to give a gentle, intuitive and non-technical introduction to objective cognitive constructivism, the epistemological framework developed to work in tandem with the $e$-value and its inference procedures. 
 %
Nowadays, $p$-values are usually presented embedded in an epistemological framework based on Popperian falsificationism, while Bayes Factors are usually presented in the context of stochastic decision theory (optimal gambling strategy). 
Stern and Pereira (2014) analyze the contrasting mathematical properties of each of the three aforementioned significance measures and the consequential differences between corresponding statistical inference procedures.   
Stern (2011a,b) analyze the contrasting characteristics of the corresponding epistemological frameworks, developed to harbor and support their specific inference methods. 
   
The objective cognitive constructivism epistemological framework has its origins in cognitive constructivism as developed by 
Maturana and Varela (1980), Foerster (2003), Foerster and Segal (2001), Krohn and K\"{u}ppers (1990), Zeleny (1980), and many others.       
Nevertheless, the objective character of the former sets it  apart from its predecessors.  
Stern (2007a,b, 2008, 2014) give  detailed analyses of the similarities and differences between these frameworks. 
Stern (2015) considers some frequent questions concerning the similarities and differences between objective cognitive constructivism and the epistemological framework developed by the philosopher Willard van Orman Quine. 

Statistical textbooks tend to de-emphasize epistemological foundations and play down the philosophical commitments made by each statistical school, presenting theory and methods in the most pragmatical way possible. 
Nevertheless, the development of each of the aforementioned statistical schools and their programs was strongly influenced or even guided the these philosophical commitments. 
Stern (2018) analyzes the work of Karl Pearson (1857-1936), founder of the frequentist school, and the importance of his philosophical commitment to 
inverted-Spinozism,  a form of transcendental idealism, to the development of the theory, methods and language of classical statistics.  
The same article analyzes the objections of K.Pearson to the epistemological views of Paul Volkmann (1856-1938), who presents his ideas based on the metaphor of \textit{the scientific system as a vaulted arch}. 
Furthermore, Stern (2018) argues that Volkmann's vaulted arch metaphor could be seen as a distant forerunner of cognitive constructivism. 
A somewhat related metaphor, \textit{the Arch of Knowledge}, is developed by David Oldroyd (1986), who uses this metaphor as his guideline to conduct a  systematic study of history, philosophy and methodology of science.  
The $e$-value and objective cognitive constructivism research program has a long way to go until a similar wide-range systematic comparison with alternative epistemological frameworks is available.  
Nevertheless, the first steps in this long journey have already been taken, focusing on the immediate vicinity of epistemological frameworks most relevant to modern statistical inference.

\section{Final Remarks} 

Trying to understand our quest for knowledge, specially in the realm of exact sciences, this article explored two main metaphors, namely, \textit{Science as provider of sharper images} of our world, and von Foerster's metaphor of \textit{Objects as tokens for eigen-behaviors}. 
We tried to show how these metaphors can be helpful in building an integrated and coherent perspective of the development of a scientific discipline, and in the  understanding of its ordinary activities.  
However, as it is often the case when a new metaphor is presented, some inherent limitations of metaphorical argumentation in general, and those we used in particular, are brought to light and should be examined. 
We shall also consider the possibility of extending the application of the epistemological framework developed in his article to  other areas, like human sciences. 
Those issues are discussed in the next subsections.

\subsubsection*{Metaphoric and Polysemic Objective Ontologies}

As shown in Sections 6, some key objects of the standard ontology of optical science directly correspond to elements in the system of eigen-functions used to describe the eigen-solutions of a vibrating disc plate, for example, the ideal forms of lenses and mirrors, and their classical aberrations.     
As shown in Stern (2008, 2014, 2017a), some key objects of the standard ontology of musical science directly correspond to elements in the system of eigen-functions used to describe the eigen-solutions of a vibrating string, for example, musical notes in classical scales.    
In either case, von Foerster's metaphor -- \textit{objects are tokens for eigen-behaviors} -- can be applied quite literary. 

Other words of these ontologies stand for compositional rules and for ways to use the preceding objects to accomplish specific goals. 
For example, Keplerian (paraxial) optics give us precise prescriptions to align and collimate the distance between lenses and mirrors in a telescope in order to obtain  well-focused images, see Appendix A and previous sections. 
In the same way, musical science has precise prescriptions on how to synchronically compose separate notes into consonant or dissonant chords, and how to diachronically arrange them into melodic lines.    
These compositional rules and their applications also express the four essential properties of eigen-solutions, namely, stability, precision, separation and composition.  
Nevertheless, the correspondence between these words and underlying systems of eigen-solutions is more remote or indirect. 
Hence, von Foerster's  metaphor can only be applied in a more abstract or figurative sense. 
Even farther removed is the connection of the underlying eigen-solutions with the variety of means and methods (and the words and language used to describe them) used by artisans and technicians in the craft of building optical or musical instruments so that they perform as expected, that is, so that they can be used to play the fine tuned music we want to hear or to render the sharp images we want to see.

Furthermore, in historical development optical instruments and their manufacturing techniques, generations of artisans have acquired practical knowledge about objects in those ontologies long before the mathematics used to formally describe a system of underlying eigen-functions was available, see King (1955) and the bibliography therein.   
In fact, the other way around, it is possible to see such mathematical formalisms as languages developed to describe, in the best way we currently can, some peculiar patterns that emerge in our ways of recurrent interaction, patterns deemed peculiar for exhibiting the aforementioned essential properties of eigen-solutions. 
In this regard, see Stern (2011a) for a discussion of  \'{A}rp\'{a}d Szab\'{o} and Imre Lakatos' views of mathematics as a quasi-empirical science.   
           
The observations in the last paragraph make us contemplate, once again, the circular, recursive and, at least to some degree,  polysemic and figurative nature of the metaphors and associated concepts used in the Objective Cognitive Constructivism epistemological framework. 
This situation may be accompanied by sensations ranging from slightly discomfort to lightheadedness or strong vertigo.   
Hence the question: 
Is this situation unavoidable? 

Would it not be possible to find a steady vantage point or an absolute reference frame from which to observe the evolution of a scientific discipline, or even to get a snapshot of its current state of affairs, in a way that dispenses intrinsically circular definitions, avoids reference to recursive processes, and evades the use of polysemic concepts or metaphorical language? 
According to the constructivist framework, the answer is in the negative. Any attempt to secure  such a privileged position would not only be futile, but essentially misconceived. 
Nevertheless, if the former task is hopeless, the general situation is promising for, as stated in one of the opening quotations of this article:         

\begin{quotation}     
	\textit{You can dance in a hurricane, but only if you are standing in the eye!}    
\end{quotation}

From the perspective of Objective Cognitive Constructivism, the best available vantage point is the eye of the hurricane, in analogy to the central point of the production diagrams depicted at Figures 9 and 10.  
Paradoxically, the very center of the spinning whirlwind is the quiet place where recursive eigen-solutions of those production processes can emerge and be sustained. 
      
Furthermore, as discussed at Sections 4 and 7, each one of the aforementioned production contexts is equipped with well-defined quantitative measures specifically designed to evaluate how well  
emerging eigen-solutions manifest their required essential properties.  
In this way, such measures act as technical touch-stones tailor-made to the task of accessing the objectivity of those entities and corresponding terms in the ontology in which they are represented.  
Such a steadfast ontology gives us a firm grip on the world, allows us to 
``carve nature at its joints''  or, at least, according to the Objective Congnitive Constructivism epistemological framework, provides the best anchors to reality that we can possibly have.   
 %%% 
 See Palma (2016) and Stern (2007b, 2011a and 2014) for further discussion of some topics briefly presented in this subsection.   
    
\subsubsection*{Eigen-Solutions and Human Sciences} 

von Foerster's metaphor -- \textit{Objects are tokens for eigen-behaviors} -- stresses the role of an interacting agent.  
In this context, recognizable objects (and their names) stand for invariant patterns that emerge in the agent's recurrent forms of interaction with his environment.  
In the case of economy, sociology, political and legal studies, psychology, and other human sciences, the agents responsible for those interactions are (in different degrees) conscious or self-aware of their role in the process, and organized in complex patterns of interaction, introducing additional and interesting aspects to be considered.  
 
Let us  examine, as a familiar example, prices of commodity goods and related assets in an economic system. 
Such prices are nominated in a currency that, in turn, can be denominated by money, like the coins in Figure 17. 
All these coins have circulated over a common geographic area in a time period spanning less then a century.      
Among these, the Euro (left, issued in 2014, featuring Galileo) is considered a strong currency; the East-German Mark (center, 1971, featuring Kepler), a weaker one; and the \textit{notgeld} or emergency money issued by the city of Halle at the interbellum hyper-inflation period (right, 1920, displaying moon and stars), the weakest. 
Why this hierarchy? 
Because of how well the eigen-values they denominate (prices in the respective economies) manifest the essential properties of an eigen-solution,  
see Debreu (1972) and Ingrao and Israel (1990) for further details. 
Cerezetti and Stern (2012) consider how specific measures of the precision, stability, and effectiveness of composition rules for prices of fundamental commodities and derivative contracts in financial markets can be used to guide either speculative trading or regulatory intervention in the same markets.   

\begin{figure}[bt] 
	%\mbox{} \vspace{0mm} \mbox{}     
	
	\centerline{  %trim=l b r t 
		\includegraphics[width=0.3\linewidth, 
		trim=0mm 0mm 0mm 0mm, clip]{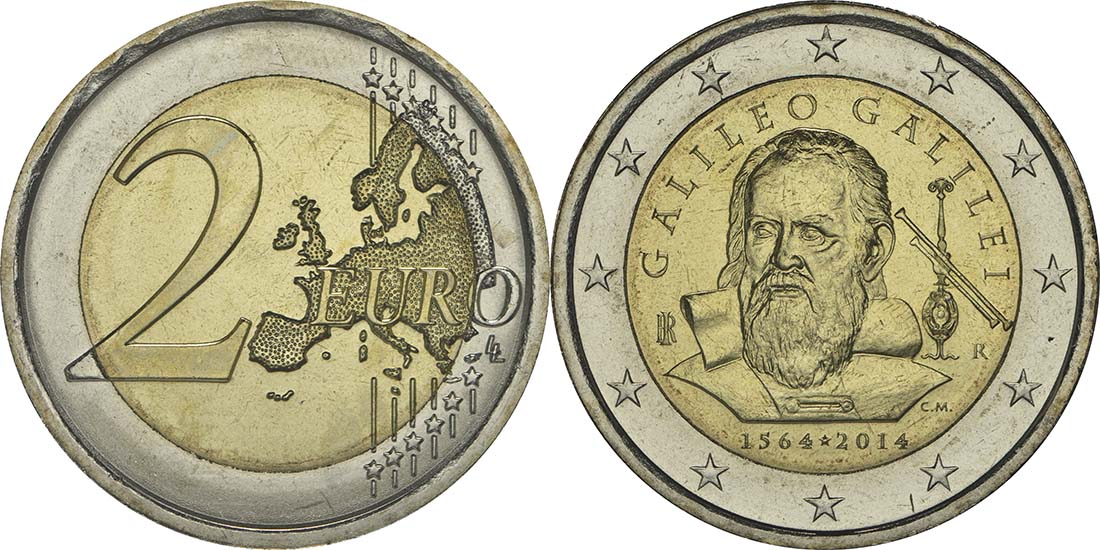}
		\mbox{}  
		\includegraphics[width=0.3\linewidth, 
		trim=0mm 0mm 0mm 0mm, clip]{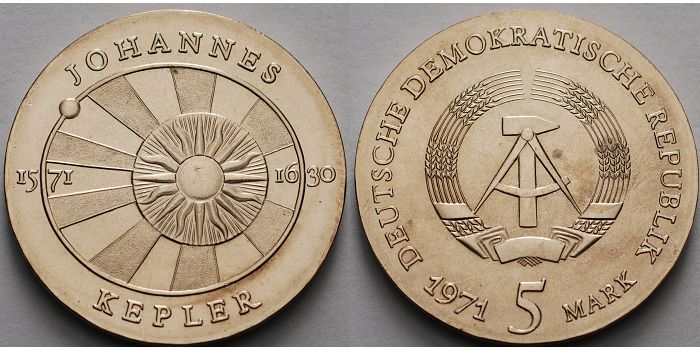} 
		\mbox{}    
		\includegraphics[width=0.3\linewidth, 
		trim=0mm 2.5mm 1mm 0mm, clip]{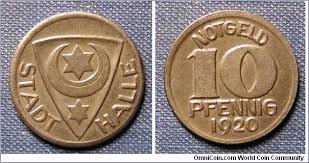}      
	}%centerline 
	
	\caption{Stronger/ weaker tokens for sharper/ blurrier eigen-values}  
	
\end{figure}

Stern (2018a) makes some incursions exploring applications to human sciences of the epistemological framework developed in this article in the area of empirical legal studies; in the near future we intend to pursue similar lines of research. 
We are particularly interested in the analysis and acceptability of causal explanations based on Niels Bohr complementarity principle in the context of economics, legal studies and psychology. 
    
\subsubsection*{Historiographical and Pedagogical Disclaimers}

In this article I made use of some narratives concerning the historical development of science and technology, and I have also described some pedagogical materials used for teaching science to children and  adolescents. 
Nevertheless, I am not an historian, and my teaching experience is limited to the university. 
Therefore, the use I made of the  aforementioned examples should be taken with a grain of salt:  
Regarding historical narratives, they often reflect common knowledge, and are not intended as position statements in the fields of historiography or history of science; see Kuukkanen (2012), Martins (2001), Sauer and Scholl (2012).  
Regarding pedagogical materials, they belong to the long-gone time of my youth, the way I describe them reflects mostly my own experiences, and these are not intended as position statements in the field.     
Development of pedagogical materials appropriate for classroom requires an holistic approach and coordinated team effort. 
For good examples of well-known pre-university texts  appropriate to the discussions in this article, see Allen el al. (1961), Batschelet (1975)*, Beck et al. (1965), Benfey et al. (1964), Elliott et al. (1972), Finetti (1957)*, Flanders and Price (1973), Flanders et al. (1973,\,1974)*, Fuller et al. (1978)*, Haber-Schaim et al. (1976), Malm et al. (1963), Pimentel et al. (1963) and Welch et al. (1968). 
As an adolescent, I used most of these books myself in Brazil at the late 70s, in preparation for the university admission exams' (vestibular) requirements of that time, see FUVEST (1977);  
Books marked by an asterisk extend somewhat beyond that scope, while 
Ayres (1958), Kindle (1950),  Rich (1963), and Schaum (1961,\,1966) provide rehearsing exercises.       
Finally, effective pedagogical projects should take into account and be adapted to the cultural background and social environment of the intended students.

\subsubsection*{Acknowledgements and Funding}    
     
I am grateful to Prof. Ac\'{a}cio Riberi (1924-1998), a blind man who, at the S\~{a}o Paulo municipal planetarium, taught me how to find my bearings by looking at the sky.     
I am thankful to Maria Julieta Sebastiani Ormastroni and her dedicated team for their contribution in developing and making available to the children of my generation in Brazil the wonderful didactic materials developed at  
% FUNBEC (Funda\c{c}\~{a}o Brasileira para o Desenvolvimento do Ensino de Ci\^{e}ncias) and    
  FUNBEC -- the Brazilian Foundation for the Development of Science Teaching, and    
% IBECC (Instituto Brasileiro de Educa\c{c}\~{a}o, Ci\^{e}ncia e Cultura) from 
  IBECC -- the Brazilian Institute for Education Science and Culture, sponsored by  
% UNESCO (Organiza\c{c}\~{a}o  das Na\c{c}\~{o}es Unidas para a Educa\c{c}\~{a}o, Ci\^{e}ncia e Cultura). 
  UNESCO -- the United Nations Educational, Scientific and Cultural Organization. 
  %%%           

I appreciate the support and constructive comments received from the organizers and participants of the  19th EBL, the Brazilian Logic Conference held on 6-10/05/19 at Jo\~{a}o Pessoa, Paraiba; and the 11th Principia International Symposium, held on 19-22/08/19 at Florian\'{o}polis, Santa Catarina.  
I am grateful for additional comments received from Rafael and Deborah Bassi Stern and anonymous referees. 

Some figures used to illustrate this article were obtained from the private
collections of Heinz Stern (currency notes,
coins and postage stamps), some were downloaded from public sites where no specific authorship information was available, and some figures were prepared for this article with the help of Alex
Freitas.  
   
This work was supported by IME-USP -- the Institute of Mathematics and Statistics of the University of S\~{a}o Paulo. 
This work was also partially supported by CNPq -- the Brazilian National Counsel of Technological and Scientific Development, grant PQ 307648/2018-4; and 
FAPESP -- the State of S\~{a}o Paulo Research Foundation, grants CEPID-CeMEAI 2013/07375-0, and CEPID-Shell-RCGI 2014/50279-4. 

 %\pagebreak 

\section*{References} 

 \renewcommand{\baselinestretch}{0.96}
 \parskip 0.75mm
 \begin{small} 
 %\begin{footnotesize} 
 %\begin{scriptsize}     %\rr   
  
\mbox{} \vspace{-4mm} \mbox{} 
  
\rr Aboufadel, Edward; Schlicker, Steven (1999).  
\textit{Discovering Wavelets}. NY: Wiley-Interscience.  	
	
\rr  Abrahams, Peter (1994). \textit{The Testing of Telescope Optics in Historic Times}. Presentation at the Convention of the Antique Telescope Society. Downloaded from: \\ 
\textit{http://www.europa.com/$\sim$telscope/histtest.txt}   
      
%\rr Ackerberg-Hastings, Amy (2015). Early Modern Computation on Sectors. p.51-61 in Zack, Maria; Landry, Elaine ed. \textit{Research in History and Philosophy of Mathematics: The CSHPM 2014 Annual Meeting}. Birkh\"{a}user, Heidelberg. 
 
\rr  Aczel,  Peter (1988). \textit{Non-Well-Founded Sets}. CSLI, Stanford, CA: Stanford University. 
   
%\rr Akivis, Maks Aizikovich; Goldberg, Vladislav Viktorovich (1977). \textit{An Introduction to Linear Algebra and Tensors}. Dover, New York. 

\rr Akman, Varol; Pakkan, M\"{u}jdat (1996). Nonstandard Set Theories and Information Management. \textit{Journal of
Intelligent Information Systems}, 6, 1, 5-31. 

\rr Aksnes, Kaare (2010). Navigation, world mapping and astrometry with Galileo's moons. 
\textit{Proceedings of the International Astronomical Union}, IAU Symposium, V.269, 42-48. 

\rr Allen, Frank B.; Douglas, E.C.; Richmond D.E.; Rickart, C.E.; Swain, H.; Walker, R.J. (1961). Intermediate Mathematics, parts 1, 2 and 3, revised ed. SMSG - School Mathematics Study Group.  New Haven: Yale University Press.   

% \rr Arrieta, Alejandro Barredo; D\'{\i}az-Rodr\'{\i}guez, Natalia; DelSer, Javier; Bennetot, Adrien; Tabik, Siham; Barbado, Alberto; Garcia, Salvador;  Gil-Lopez, Sergio; Molina, Daniel; Benjamin, Richard; Chatila, Raja; Herrera, Francisco (2019). 
% Explainable Artificial Intelligence (XAI): Concepts, Taxonomies, Opportunities and Challenges toward Responsible AI. 
%\textit{arXiv:1910.10045v1}. 

%\rr Arvacheh, E. M.;  Tizhoosh, H. R. (2005). Pattern analysis using Zernike moments. In Proceedings of the 2005 IEEE Instrumentation and Measurement Technology Conference, v.2, pp. 1574-1578. 

\rr Ayres, Frank (1958). \textit{Theory and Problems of First Year College Mathematics}. NY: Schaum.    

%\rr Bahm, Archie J. 1944. Teleological Arguments. \textit{Scientific Monthly}, 58, 377-382. 

\rr Barwise, Kenneth Jon; Etchemendy, John (1987). \textit{The Liar: An Essay on Truth and Circularity}. Oxford Univ. Press. 

\rr Barwise, Kenneth Jon; Moss, Larry (1996). \textit{Vicious Circles. On the Mathematics of Non-Wellfounded Phenomena}. Stanford, USA: CSLI.
 
\rr Batschelet, Edward (1975). \textit{Introduction to Mathematics for Life Scientists}. Berlin: Springer.
 
\rr Beck, Alexander et al. (1965). Calculus, parts 1, 2 and 3, revised ed. SMSG - School Mathematics Study Group. Stanford University.     
     
%\rr Bedini, Silvio A. (1966). Lens Making for Scientific Instrumentation in the Seventeenth Century. \textit{Applid Optics}, 5, 5, 687-694. 

%\rr Bedini, Silvio A. (1967). The Makers of Galileo's Scientific Instruments. Atti del Symposium Internazionale di Storia, Metodologia, Logica e Filosofia della Scienza; Galileo nella Storia e nella Filosofia della Scienza, 89-115. Vinci: Gruppo Italiano di Storia delle Scienze. 

%\rr Bedini, Silvio A. (1967). An Early Optical Lens-Grinding Lathe. \textit{Technology and Culture}, 8, 1, 74-77. 

\rr Bellosta, H\'{e}l\`{e}ne (2002). Essay-Review: 
Burning Instruments: From Diocles to Ibn Sahl. 
\textit{Arabic Sciences and Philosophy}, 23, 285-303.

\rr Benade, Arthur H. (1960). \textit{Horns, Strings and Harmony}. Garden City, NY: Anchor.  

\rr Benfey O.T.; Clapp, L.B.; Derose, J.V.; Hunter, W.E.; Livermore, A.H.; Neidig, H.A.; Scheer, M.H.; Scott, E.S.; Strong, L.E.; Wilson, M.K. (1964). \textit{Chemical Systems: CBA - Chemical Bond Approach Project}. NY: McGraw-Hill.

\rr Bernieri Enrico (2012). Learning from Galileo's 
 Errors. \textit{Journal of the British Astronomical Association}, 122, 3,  169-172.

\rr Biagioli, Mario (1993). \textit{Galileo, Courtier: The Practice of Science in the Culture of Absolutism}. Univ.\,of Chicago Press. 
%p.35. 

\rr Biagioli, Mario (2006). \textit{Galileo's Instruments of Credit: Telescopes, Images, Secrecy}. Univ.\,of Chicago Press. 
%Ch.2 p.77-134. 

\rr Boorstin, Daniel J. (1983). \textit{The Discoverers}. NY: Random House.  

\rr Borges, Wagner; Stern, Julio Michael (2007). The Rules of Logic Composition for the Bayesian Epistemic E-Values. 
\textit{Logic Journal of the IGPL}, 15(5/6), 401-420.

\rr Borrelli, Arianna; Hon, Giora; Zik, Yaakov eds. (2017). \textit{The Optics of Giambattista Della Porta (ca. 1535–1615): A Reassessment}. Cham: Springer.  

\rr Brecht, Bertolt (1955). Life of Galileo (Leben des Galilei). In  \textit{Bertolt Brecht: Plays, Poetry and Prose}. London: Methuen, 1980. 

% \rr Brier, S{\o}ren (2005). ``The Construction of Information and Communication: A Cybersemiotic Reentry into Heinz von Foerster's Metaphysical Construction of Second-Order Cybernetics''. \textit{Semiotica}, 154(1/4), 355-399. 

% \rr P.Brunet (1938). \textit{\'{E}tude Historique sur le Principe de la Moindre Action.} Paris: Hermann.

%\rr Bucciantini, Massimo; Camerota, Michele; Giudice, Franco (2015).  \textit{Galileo’s Telescope: A European History}. Harvard University Press, Cambridge, Massachusetts. 

%\rr Burnett, Graham D. (2005). \textit{Descartes and the Hyperbolic Quest: Lens Making Machines and their Significance in the Seventeenth Century}. American Philosophical Society, Philadelphia.

\rr Butkov, Eugene (1968). \textit{Mathematical Physics}. Reading, MA: Addison-Wesley. 

\rr Cerezetti, Fernando Valvano; Stern, Julio Michael (2012). Non-arbitrage in Financial Markets: A Bayesian Approach for Verification. 
\textit{American Institute of Physics Conference Proceedings}, 
1490, 87-96. 
 
\rr Civita, Victor ed. (1972). \textit{Os Cientistas}. S\~{a}o Paulo: Abril Cultural.   
  
\rr Courant,  Richard; Hilbert, David (1953). \textit{Methods of Mathematical Physics}. NY: Wiley. 
     
\rr  Debreu, Gerard (1972). \textit{Theory of Value: An Axiomatic Analysis of Economic Equilibrium}. New Haven: Yale University Press. 
    
\rr Dugas, Ren\'{e} (1988).  
\textit{A History of Mechanics}. NY: Dover. 
 %Ch.5 The Principle of Least Action, p.254-275

%\rr Dupr\'{e}, Seven (2008). Inside the Camera Obscura: Kepler's Experiment and Theory of Optical Imagery. \textit{Early Science and Medicine}, 13, 219-244.
%della Porta's tricks: p.241-242   

%\rr Dzhaparidze, K.O. (1986). \textit{Parameter Estimation and Hypothesis Testing in Spectral Analysis of Stationary Time Series}. NY: Springer-Verlag. 

%\rr Eastwood, Bruce Stansfield (1984). Descartes on Refraction: Scientific versus Rhetorical Method. \textit{Isis}, 75, 3, 481-502. 

%\rr Ehring, Douglas (1984). The System-Property Theory of Goal-Directed Processes. Philosophy of the Social Sciences, 14, 4, 497-504. 

%\rr A.Einstein (1950). On the Generalized Theory of Gravitation. \textit{Scientific American,} 182, 4, 13-17. Reprinted in Einstein (1950, 341-355). 

%\rr A.Einstein (1954). \textit{Ideas and Opinions}. Wings Books.  % Mein Weltbild. 

% \rr A.Einstein (1991). \textit{Autobiographical Notes: A Centennial Edition.} Open Court Publishing Company. 

\rr Elliott, Harold Andrew; Fryer Kenneth D.; Gardner James C.; Hill, Norman J. (1972). 
\textit{I- Relations, Transformations and Statistics}. 
\textit{II- Vectors, Matrices and Algebraic Structures}. 
\textit{III- Calculus, Complex Numbers and Polar Co-Ordinates}. Toronto: Holt, Richart and Winston.  

% \rr Elsgolts, Lev (1977). \textit{Differential Equations and the Calculus of Variations}. MIR, Moscow. 
   
\rr Esteves, Luis Gustavo; Izbicki, Rafael; Stern, Julio Michael; Stern, Rafael Bassi (2016). 
The logical consistency of simultaneous agnostic hypothesis tests. \textit{Entropy}, 18, 256, 1-22.  

\rr Esteves, Luis Gustavo; Izbicki, Rafael; Stern, Julio Michael; Stern, Rafael Bassi (2019). 
Pragmatic Hypotheses in the Evolution of Science. 
\textit{Entropy}, 21, 883, 1-17.  
%\textit{doi.org/10.3390/e21090883}  

\rr Favaro, Antonio (1983). \textit{Amici e corrispondenti di Galileo}. Florence: Paolo Galluzzi. 

\rr Ferguson, Kitty (2002). \textit{Tycho \& Kepler: The Unlikely Partnership That Forever Changed Our Understanding ofthe Heavens}. NY: Walker \& Company.  

\rr Fermat, Pierre de (1894). 
\textit{Oeuvres}, V.2, Correspondance. 
Gauthier-Villars, Paris, 1894. Including: 
(a) Letter of Fermat to de la Chambre of 
August 1657, LXXXVI, p.354-359. 
(b) Letter of Clerselier to Fermat of 
May 6th, 1662, CXIII, 464-472. %(p.465). 

\rr Feynman Richard Phillips; Leighton, Robert; Sands Matthew (1964). \textit{The Feynman Lectures on Physics}.   
Boston: Addison-Wesley. %v.II, Chapter 19. 
 
\rr Feynman, Richard Phillips; Hibbs, Albert Roach (1965). \textit{Quantum Mechanics and Path Integrals}.  NY: McGraw-Hill. 

\rr Feynman, Richard Phillips (1985). \textit{The Character of Physical Law}. Cambridge, MA: MIT Press.

\rr Finetti, Bruno de (1957). \textit{Matematica Logico-Intuitiva}. Roma: Edizioni Cremonese.  

\rr Finocchiaro, Maurice A. (1989). \textit{The Galileo Affair: A Documentary History}. Berkeley: University of California Press. 
  
\rr Flanders, Harley R.;  Price, Justin J. (1973). 
\textit{Introductory College Mathematics with Linear Algebra and Finite Mathematics}. NY: Academic Press. 
  
\rr Flanders, Harley R.; Korfhage, Robert R.; Price, Justin J. (1973) \textit{A First Course in Calculus with Analytic Geometry}. NY: Academic Press. 

\rr Flanders, Harley R.; Korfhage, Robert R.; Price, Justin J. (1974) \textit{A Second Course in Calculus with Analytic Geometry}. NY: Academic Press. 
      
\rr Foerster, Heinz von; Segal, Lynn (2001). \textit{The Dream of Reality: Heinz von Foerster's Constructivism}. NY: Springer-Verlag. 
  
\rr Foerster, Heinz von (2003). \textit{Understanding Understanding: Essays on Cybernetics and Cognition}. NY: Springer.   

\rr  Foucault, L\'{e}on (1858). Description des procedees employes pour reconnaitre la configuration des surfaces optiques. \textit{Comptes rendus hebdomadaires des s\'{e}ances de l'Acad\'{e}mie des Sciences}, 47, 958-959.

\rr Foucault, L\'{e}on (1859). M\'{e}moire sur la construction des t\'{e}lescopes en verre argent\'{e}. 
\textit{Annales de l'Observatoire imp\'{e}riale de Paris}, 5, 197-237.  
  
\rr Fracastoro, Girolamo (1538). \textit{Homiocentrica. 
Eiusdem de causis criticorum dierum per quae in nobis sunt}. Venice, 18v, Sectio II, cap. 8.   
%and 58r (Sectio III, cap. 23).

\rr  French, Anthony Philip (1971). \textit{Vibrations and Waves}. The M.I.T. Introductory Physics Series. NY: W.W.Norton.

\rr Fr\"{o}hlich, Wilhelm (1923). \textit{Anleitung zum Gebrauch des Kosmos-Baukasten Optik: Versuche aus der Lehre vom Licht}. Stuttgart: Kosmos Gesellschaft der Naturfreunde.  
%Franckh'sche Verlagshandlung, 

\rr Fuller, Harold Q.; Fuller, Richard M.; Fuller, Robert G. (1978). \textit{Physics: Including Human Applications}. NY: Harper and Row.  

\rr FUVEST (1977). {Manual do Candidato: Vestibular 1977}. S\~{a}o Paulo: FUVEST.  

\rr Galilei, Galileo (1900). \textit{Le Opere di Galileo Galilei}. Edited by Antonio Favaro. Edizione Nazionale. 20 vols. Firenze: G. Barb\`{e}ra, 1892-1904. Vol. X, Carteggio 1574--1642.

%\rr Galilei, Galileo (1606). \textit{Le Operazioni del Compasso Geometrico et Militare}; Padova. Translated as \textit{Operations of the Geometric and Military Compass}. The Smithsonian Institution Press, Washington, 1978.
  
\rr Galilei, Galileo (1610). \textit{Sidereos Nuncios}; 
Venice. Translated as \textit{The Sidereal Messenger}. 
Univ.\,of Chicago Press, 1989. 
         
\rr Garcia, Manuel Valentim Pera; Humes, Carlos; Stern, Julio Michael (2003). 
Generalized Line Criterion for Gauss-Seidel Method. 
\textit{Computatational and Applied Mathematics}, 22, 91-97.          

\rr Genberg, Victor; Doyle, Keith; Michels, Gregory (2004). Optical interface for MSC. Nastran. 
\textit{MSC Conference Proc.}, VPD04-31. 
         
\rr Gerrard, Anthony; Burch, James M. (1975). 
\textit{Introduction to Matrix Methods in Optics}.  
London: John Wiley.     

\rr Goldstein, Herbert  (1980). \textit{Classical Mechanics}, 2nd ed. Reading: Addison-Wesley.  
		
\rr Goldstine, Herman H. (1980). \textit{A History of the Calculus of Variations from the 17th through the 19th Century}. NY: Springer-Verlag. 
   
% \rr Guckenheimer, John; Holmes, Philip (1983). 
%\textit{Nonlinear Oscillations, Dynamical Systems, and 
%Bifurcations of Vector Fields}. Springer-Verlag, New York. 
% Sec. 1.4 Linear and Nonlinear Maps p.19-22 
% Sec. 1.5 Closed Orbits, Poincar\'{e} Maps, 
%          and Forced Oscillations.  

%\rr  Harbour, David A. (2013). \textit{Understanding 
%	 Foucault}.  Sapphire Publications. 
   
\rr Haber-Schaim, Uri; Cross, Judson B., Dodge, John E.; Walter, James A. (1976). 
\textit{PSSC Physics}, Physical Science Study Committee, 4th ed. Lexington, MA: D.C. Heath.  
     
\rr  Harbour, David A. (2001). \textit{Understanding 
Foucault: A Primer for Beginners}. ATM's Workshop -- Tips, Techniques, and Projects for the Amateur Telescope Maker. \\  
\textit{www.atm-workshop.com/foucault.html} 
       
\rr Heeffer, Albrecht (2017). Using Invariances in 
Geometrical Diagrams: Della Porta, Kepler and 
Descartes on Refraction.  Ch.7, p.145-168 in Borrelli (2017). 
  
\rr Helden, Albert van (1977). \textit{The Invention of the Telescope}. Transactions of the American Philosophical Society, 67, 4, 1-67.    
         
\rr Helden, Albert van; Dupr\'{e}, Scen; Gent, Rob van; 
Zuidervaart, Huib (2010). \textit{The Origins of the Telescope}. Amsterdam: KNAW Press. 

\rr Helrich, Carl S. (2007). Is there a Basis for Teleology in Physics?  \textit{Zygon}, 42, 1, 97-110.
  
%\rr Hentschel, Klaus (2001). Das Brechungsgesetz in der Fassung  von Snellius. \textit{Archive for History of Exact Sciences}, 55,4, 297–344. 
%Rekonstruktion seines Entdeckungspfades und eine \"{U}bersetzung 
%seines lateinischen Manuskriptes sowie erg\"{a}nzender Dokumente  

\rr Holton, Gerald James (1988). \textit{Thematic Origins of Scientific Thought: Kepler to Einstein}. 
 Harvard University Press. 

\rr Hu, Ming-Kuei (1962). Visual Pattern Recognition by Moment Invariants. \textit{IRE transactions on information theory}, 8, 2, 179-187.

\rr Ilardi, Vincent (2007). \textit{Renaissance  Vision: 
From Spectacles to Telescopes}.
Philadelphia: American Philosophical Society.  
		
\rr Ingalls, Albert G. (1953). %(1996). 
\textit{Amateur Telescope Making}, Vols. 1, 2 and 3.    
Kingsport, TN: Scientific American Inc. %Willmann-Bell.  	

\rr Ingrao, Bruna; Israel, Giorgio (1990). \textit{The Invisible Hand. Economic Equilibrium in the History of Science}. Cambridge, MA: The MIT Press.

\rr Iordanov, Borislav (2010). HyperGraphDB: A generalized graph database. \textit{LNCS}, 6185, 25-36. 

\rr Izbicki, Rafael; Esteves, Luis Gustavo (2015).  Logical consistency in simultaneous statistical test procedures. \textit{Logic Journal of the IGPL}, 23, 732–758.
    
\rr Jaffe, Bernard (1960). \textit{Michelson and the Speed of Light}. 	Anchor Books, Garden City, NY. 	
  
\rr Jarvis, Matt (2016), Waves \& Normal Modes. 
21/05/2016, retrieved from \\ 
\textit{https://www2.physics.ox.ac.uk/sites/default/files/2012-09-04/fullnotes\_pdf\_15169.pdf}  

%\rr Kac, Mark (1966). Can One Hear the Shape of a Drum? \textit{American Mathematical Monthly}, 73, 4, 2, 1-23.

%\rr Katzav, J. (2004). Dispositions and the Principle of Least Action. \textit{Analysis}, 64.3, 206-214. 

\rr Kelley, Carl Timothy (1987a). 
\textit{Iterative Methods for Linear and Nonlinear Equations}. Philadelphia: Society for Industrial Mathematics. 

\rr Kelley, Carl Timothy (1987b). 
\textit{Iterative Methods for Optimization}.  
Philadelphia: Society for Industrial Mathematics.

\rr Kepler, Johannes (1604, 2000). \textit{Ad Vitellionem Paralipomena, quibus Astronomiae pars Optica}. 
%In his \textit{Gesammelte Werke} (), 2, 181-xxx.
Frankfurt: Claudium Marnium \& Ioannis Aubrii.  
Translated by  W.H. Donahue as 
\textit{Optics: Paralipomena to Witelo and Optical 
Part of Astronomy}. Santa Fe: Lion Press.   
% William Haeredes 
		
\rr Kepler, Johannes (1611). \textit{Dioptrice}. 
In his \textit{Gesammelte Werke} (1937), 4, 355-414.   

\rr Kepler, Johannes (1610, 1965). 
\textit{Dissertatio, cum Nuncio Sidereo, nuper ad mortales 
misso a Galilaeo Galilaeo, Mathematico Patauino}. 
translated by Edward Rosen as 
\textit{Conversation with Galileo's Sideral Messenger}.   
NY: Johnson Reprint Co. 

% \rr Katzav, J. (2005). Ellis on the limitations of dispositionalism. \textit{Analysis}, 65, 92-94. 

\rr Khan, Sameen Ahmed (2015). 
Medieval Islamic achievements in optics. 
\textit{Il Nuovo Saggiatore}, 31, 1, 36-45.  
%%%% Boa exsplicacao! 

\rr Khotanzad, Alireza; Hong, Yaw Hua (1990). Invariant Image Recognition by Zernike Moments. \textit{IEEE Transactions on Pattern Analysis and Machine Intelligence} 12, 5, 489-497.

\rr Kindle, Joseph H. (1950). \textit{Theory and Problems of Plane and Solid Analytic Geometry}. NY: Schaum.    
   
\rr King, Henry C. (1955). \textit{History of the Telescope}. NY: Dover. 
        
\rr  Kleinfeld, David; Tsai, Philbert (2004).  
\textit{An introduction to basic optical design: Matrix techniques through scanning microscopy}. 
 Univ.\,of California at San Diego, 13/04/2004. \\ 
\textit{https://neurophysics.ucsd.edu/courses/physics\_173\_273/Phys\_173\_optics.pdf} %\\ 
 %\textit{https://neurophysics.ucsd.edu/courses/physics\_173\_273/optics\_matrix\_methods\_lecture\_notes.pdf}
   
\rr Koestler, Arthur (1960). \textit{The Watershed: A Biography of Johannes Kepler}. Garden City, NY: Anchor Books. 	
	   
%\rr Koppany, Bob (2010). Magnifiers. \textit{Journal of the Oughtred Society}, 19, 2, 28-31.   

\rr  Krasnov, Mihail Leontievich; Makarenko, Grigoi Ivanovich;  Kiseliov Aleksandr Ivanovich (1973, 1984). \textit{C\'{a}lculo Variacional}. Moskow: MIR. 

\rr  Krohn, Wolfgang; K\"{u}ppers G\"{u}nter (1990).  
The Selforganization of Science: Outline of a Theoretical Model. p.208-222 in 
Krohn, Wolfgang; K\"{u}ppers, G\"{u}nter; Nowotny, Helga; \textit{Selforganization: Portrait of a Scientific Revolution}. Dordrecht: Kluwer.   
    
\rr Kuukkanen, Jouni-Matti (2012). The missing narrativist turn in the historiography of science. 
\textit{History and theory}, 51.3 (2012): 340-363.
  
\rr Lakshminarayanana, Vasudevan; Flecka, Andre (2011). Zernike Polynomials: A Guide. 
\textit{Journal of Modern Optics}, 58, 7, 545-561.  
  
%\rr Larmer, J. W.; Goldstein, E. (1966). Some Comments upon Current Optical Shop Practices. \textit{Applied Optics}, 5, 5, 677-685.
     	
%\rr Lattis, James M. (1994). \textit{Between Copernicus and Galileo: Christoph Clavius and the Collapse of Ptolemaic Cosmology}. University of Chicago Press.			

\rr Leech, John Watson (1963). \textit{Classical Mechanics}.  London: Methuen. 
			
%\rr  Lef\`{e}vre, Wolfgang (2007). \textit{Inside the Camera Obscura: Optics and Art under the Spell of the Projected Image}. Preprint 333, Max-Planck Institut f\"{u} Wissenschaftsgeschichte,
% Max Planck Institute for the History of Science

\rr Lemons, Don S. (1997). \textit{Perfect Form: Variational Principles, Methods and Applications to Elementary Physics}. Princeton Univ. Press. 
   
%\rr Li, Ta-Hsin (2013).  \textit{Time Series with Mixed Spectra}. Boca Raton: CRC Press. 	
	
%\rr D.Loemker (1969). \textit{G.W.Leibniz Philosophical Papers and Letters.}  Reidel.		

%\rr Allan Mackintosh, G. Kenneth Hawkings, Ferdinand I. Baar (1986). \textit{Advanced Telescope Making: Optics (v1), Mechanical (v2)}. Richmond, VA: Willmann-Bell. 

\rr Madruga, Maria Regina; Esteves, Luis Gustavo;  Wechsler, Sergio (2001). On the Bayesianity of Pereira-Stern Tests. \textit{Test}, 10, 291-299.

%\rr Mahajan, Virendra N. (2010). Orthonormal Polynomials in Wavefront Analysis. In \textit{Handbook of Optics}, 3rd ed. Vol 2, Part 3, Chapter 11. McGraw-Hill, New York.  

\rr Malet, Antoni (2003). Kepler and the Telescope. \textit{Annals of Science}, 60, 2, 107-136.  

\rr Malm, Lloyd E.; Davis, Joseph E.; Nicholson, Margart (1963). 
\textit{Laboratory Manual for Chemistry: An Experimental Science}, Chemical Education Material Study. San Francisco: W.H. Freeman. 

%\rr Manzini, Carlo Antonio  (1660). \textit{L'occhiale all'occhio. Dioptrica pratica}. l'Herede del Benacci, Bologna. 

\rr Marion, Jerry B. (1970). \textit{Classical Dynamics of Particles and Systems}, 2nd ec. NY: Academic Press. 

\rr Martens, Rhonda (2000). \textit{Kepler's Philosophy and the New Astronomy}. Princeton, NJ: Princeton Univ. Press. 

\rr Martins, Roberto de Andrade (2001). Como n\~{a}o escrever sobre hist\'{o}ria da F\'{\i}sica: Um manifesto historiogr\'{a}fico. \textit{Revista Brasileira de F\'{\i}sica}, 23, 1, 113-129.   

\rr Maturana, Humberto (1997). \textit{La Objectividad: Um argumento para Obligar}. Santiago de Chile: Dolmen.  
		
%\rr Maturana, Humberto;  Varela, Francisco Javier (1980). \textit{Autopoiesis and Cognition: The Realization of the Living}.  D. Reidel Publishing, Dordrecht.

\rr McDonough, Jeffrey K. (2009). Leibniz on Natural Teleology and the Laws of Optics. \textit{Philosophy and Phenomenological Research}, 78, 3, 505-544. 

\rr McDonough, Jeffrey K. (2015). Descartes' Optics. 
p. 550-559 in Nolan, Lawrence (2015). \textit{The Cambridge Descartes Lexicon}. Cambridge: Cambridge University Press. 

\rr Mihas, Pavlos (2008). Developing Ideas of Refraction, Lenses and Rainbow Through the Use of Historical Resources. 
\textit{Science \& Education}, 17, 7, 751-777.

%\rr Molesini, Giuseppe (2010). Telescope Lens-Making in the 17 th Century: The Legacy of Vangelista Torricelli. \textit{Optics and Photonics News}, 21, 4, 26-31.
  
\rr Molesini, Giuseppe (2011). Early advances on rays and refraction: a review through selected illustrations. \textit{Optical Engineering}, 50, , 121704, 1-6.  

%\rr Morinelli, Tony Devaney (2016).  Reason and Doctrine: Time for Christians to Rethink What They Believe. NY: Algora. 

\rr Morrison, Philip; Morrison, Phylis (1987). \textit{The Ring of Truth: Inquiry Into How We Know What We Know}. Random House. 
Video \textit{Lenses and Telescopes} from the homonymous PBS series downloaded in 3 parts from: \  
\textit{www.youtube.com/watch?v=zqQs6LauvUc} \ ; \\ 
\textit{www.youtube.com/watch?v=rlVt7EPoiaM} \ ; \   
\textit{www.youtube.com/watch?v=9HnpHbgr2hU}

\rr Muir, Edward (2007). \textit{The Culture Wars of the Late Renaissance: Skeptics, Libertines, and Opera}. Cambridge, MA: Hardvard Univ. 

%\rr I.L. Muntean (2006). \textit{Beyond Mechanics: 
%Principle of Least Action in Maupertuis and Euler.} 
%On line doc., University of California at San Diego.   

\rr Nagel, Ernest  (1979). Teleology Revisited. 
in \textit{Teleology Revisited and Other Essays in the Philosophy of Science}. NY: Columbia Univ. Press. 

\rr  Nakano, Hideya; Martins, Roberto Andrade; Krasylchik, Myriam ed. (1972). \textit{Os Cientistas}, v.19 - Descartes: A Trajet\'{o}ria dos Raios Luminosos. S\~{a}o Paulo: FUNBEC/ Abril Cultural. 

%\rr  Nietzsche, Friedrich (1873). \"{U}ber L\"{u}ge und Wahrheit im aussermoralischen Sinn. Werke, v.3, p.309-322, Munich: Hanser, 1954.  
%Translated as: Truth and Lies in a Non-Moral Sense. p.79-101 in Breazeale, D. (1979). Philosophy and Truth. Atlantic Highlands, NJ: Humanities Press. 
    
\rr Oldroyd, David (1986). \textit{The Arch of Knowledge: An Introductory Study of the History of the Philosophy and Methodology of Science}. Kensington: New South Wales University Press. %Translated as \textit{Storia della filosofia della scienza}, 1989, Milano: Il Saggiatore.   
    
\rr Palma, H\'{e}ctor (2016). \textit{Ciencia y Met\'{a}foras: Cr\'{\i}tica de una Raz\'{o}n Incestuosa}. Buenos Aires: Prometeo. 

\rr  Pakkan, M\"{u}jdat; Akman, Varol (1995). Hypersolver: A graphical tool for commonsense set theory. \textit{Information
Sciences}, 85, 1-3, 43-61. 
    
\rr Pedrotti, Frank L.; Pedrotti, Leno M.; Pedrotti, Leno S. (2017). \textit{Introduction to Optics} 
Cambridge University Press.

\rr Pereira, Carlos Alberto de Bragan\c{c}a; Stern, Julio Michael (1999). Evidence and credibility: Full bayesian significance test for precise hypotheses. 
\textit{Entropy}, 1, 99–110.

\rr Pereira, Carlos Alberto de Bragan\c{c}a; Stern, Julio Michael (2020). The e-value: A Fully Bayesian Significance Measure for
Precise Statistical Hypotheses and its Research Program. 
\textit{arXiv:2001.10577v2}
    
\rr Pereira, Carlos Alberto de Bragan\c{c}a; Stern, Julio Michael; Wechsler, Segio (2008). Can a Signicance Test be Genuinely Bayesian? \textit{Bayesian Analysis}, 3, 79–100. 
  
\rr Pereira, Mateus H. F. (2005). 
A Trajet\'{o}ria da Abril Cultural (1968-1982). 
\textit{Em Quest\~{a}o}, 11, 2, 239-258.  

%\rr Pettrofrezzo, Anthony J. (1978). \textit{Matrices and Transformations}. Dover, New York.  

\rr Pimentel, George C.; Mahn, Bruce H.; McClellan, A.H.; MacNab, Keith; Nicholson, Margaret, eds. (1963). 
\textit{Chemistry: An Experimental Science}, Chemical Education Material Study. San Francisco: W.H. Freeman. 

%\rr Planck Max (1915). Das Prinzip der kleinsten Wirkung. {\it Kultur der Gegenwart}. Also in p.25-41 of Planck (1944). 

%\rr Planck max (1944) Wege zur physikalischen Erkenntnis. Reden und Vortr\"{a}ge. Leipzig: S.Hirzel.  

%\rr Planck Max (1937). Religion and Natural Science. Also in Planck (1950).  

%\rr Planck Max (1950). Scientific Autobiography and other Papers. London: Williams and Norgate. 

% \rr Porta, Giovanni Baptista Della (1560). 
% \textit{Dei Miracoli et Maravigliosi Effetti dalla Natura 
% Prodotti.}. L. Avanzi, Venice. % 140–142.

\rr Porta, Giovanni Baptista Della (1593). 
\textit{De Refractione Optices, parte libri IX}.  
Neapoli: Ex officina Horatii Salviani, 
apud Jo. Jacobum Carlinum, \& Antonium Pacem.

\rr Porta, Giovanni Baptista Della (1658).   
On Strange Glasses -- Wherein are propounded 
Burning-glasses, and the wonderful sights 
to be seen by them.  Book XVII, p.355-381, in 
\textit{Natural Magick: In XX Bookes by John Baptist 
Porta, a Neopolitane}.  London: Thomas Young \& Samuel Speed. 
Translation from the original \textit{Magiae Naturalis sive 
de Miraculis Rerum Naturalium}, 2nd ed., 1589; 1st ed, 1558.

%\rr H.Pulte (1989).  Das Prinzip der kleinsten Wirkung und die Kraftkonzeptionen der rationalen Mechanik: Eine Untersuchung zur Grundlegungsproblemematik bei Leonhard Euler, Pierre Louis Moreau de Maupertuis und Joseph Louis Lagrage. \textit{Studia Leibnitiana}, sonderheft 19.   		
		
%\rr Rashed, Roshdi (1990). A Pioneer in Anaclastics: Ibn Sahl on Burning Mirrors and Lenses. \textit{Isis}, 81, 464-464. 

%\rr Rashed, Roshdi (1993). \textit{G\'{e}om\'{e}trie et Dioptrique au X Si\`{e}cle: Ibn Sahl, al-Quhi et Ibn al-Haytham}. Les Belles Lettres, Paris.  

\rr Rashed, Roshdi (1993). \textit{Geometry and Dioptrics in Classical Islam}. London: Al-Furqan.  

\rr Reddy, Junuthula N. (2007). \textit{Theory and analysis of elastic plates and shells}. Boca Raton: Taylor and Francis.
   
\rr Resnikoff, Howard L.;   Wells, Raymond O. (1998). 
\textit{Wavelet Analysis: The Scalable Structure of Information}, NY: Springer-Verlag. 	
	
%\rr Rijsbergen, Keith van (2004). \textit{The Geometry of Information Retrieval}. Cambridge University Press.

\rr Rich, Barnett (1963). \textit{Principles and Problems of Plane Geometry}. NY: Schaum.    

\rr Ronchi, Vasco (1970). \textit{The Nature of Light}. 
Harvard Univ. Press, Cambridge, USA. 

%\rr Rottman, Gerald (2008). \textit{The Geometry of Light: Galileo's Telescope, Kepler's Optics}. Gerald Rottman, Baltimore. 
  
\rr Sabra, Abdelhamid I. (1981) \textit{Theories of Light from Descartes to Newton}. Cambridge University Press.
  
\rr Sadun, Lorenzo Adlai (2007). \textit{Applied Linear Algebra: The Decoupling Principle}. Providence: American Mathematical Society. 
  
%\rr Saichev, Alexander I.;  Woyczy\'{n}ski, Wojbor A. (1996). \textit{Distributions in the Physical and Engineering Sciences: Distributional and Fractal Calculus, Integral Transforms and Wavelets}. Basel: Birkh\"{a}user.  
  
\rr Santillana, Giorgio de (1978). 
\textit{The Crime of Galileo}. Chicago Univ. Press. %p.9 
%\textit{El crimen de Galileo: 
%Historia del proceso inquisitorial al genio}. 
%Antonio Zamora, Buenos Aires.  %p.22        

\rr Sauer, Tilman; Scholl, Raphael (2012). \textit{The Philosophy of Historical Case Studies}. 
Boston Studies in the Philosophy and History of Science, 319. Cham: Springer. 

\rr Schaum, Daniel (1961). \textit{Theory and Problems of College Physics}. NY: Schaum.    

\rr Schaum, Daniel (1966). \textit{Theory and Problems of College Chemistry}. NY: Schaum.    

% \rr  M.Schlick (1920). Naturphilosophische Betrachtungen \"{u}ber das Kausalprintzip. {\it Die Naturw issenschaften}, 8, 461-474. Translated as, Philosophical Reflections on the Causal Principle, ch.12, p.295-321, v.1 in M.Schlick (1979).    

% \rr M.Schlick (1979). Philosophical Papers. Dordrecht: Reidel. 

\rr Shea, James H. (1998). Ole Romer, the speed of light, the apparent period of Io, the Doppler effect, and the dynamics of Earth and Jupiter. \textit{American Journal of Physics}, 66, 561-569. 

\rr Sirtori, Giovanni (1618). \textit{Telescopium: Sive ars perficiendi novum illud Galilaei visorium instrumentum ad sidera}. Frankfurt.  

%\rr Smith, A. Mark (1987). \textit{Descartes's Theory of Light and Refraction: A Discourse on Method}. Philadelphia: The American Philosophical Society. 

\rr Sommerhoff, Gerd (1969). The abstract characteristics of living systems. p.147-202 in F. E. Emery (ed.), \textit{Systems Thinking}. Baltimore: Penguin Books. 

\rr Sommerhoff, Gerd (1990).
Life, Brain and Consciousness: New Perceptions through Targeted Systems Analysis. Amsterdam: North Holland. 	

\rr Stephenson,  Bruce (1987). \textit{Kepler's Physical Astronomy}. NY: Springer-Verlag. 

\rr Stern, Julio Michael (2007a). Cognitive Constructivism, Eigen-Solutions, and Sharp Statistical Hypotheses. \textit{Cybernetics \& Human Knowing}, 14, 1, 9-36.

\rr Stern, Julio Michael  (2007b). Language and the Self-Reference Paradox. \textit{Cybernetics \& Human Knowing}, 14, 4, 71-92.

\rr Stern, Julio Michael (2008). Decoupling, Sparsity, 
Randomization, and Objective Bayesian Inference. 
\textit{Cybernetics \& Human Knowing}, 15, 2, 49-68.
  
\rr Stern, Julio Michael  (2011a). Symmetry, Invariance and Ontology in Physics and Statistics. \textit{Symmetry}, 3, 3, 611-635.

\rr Stern, Julio Michael  (2011b). Constructive Verification, Empirical Induction, and Falibilist Deduction: A Threefold Contrast. \textit{Information}, 2, 635-650. 

\rr Stern, Julio Michael  (2014). Jacob's Ladder and Scientific Ontologies. \textit{Cybernetics \& Human Knowing}, 21, 3, 9-43.

\rr Stern, Julio Michael  (2015). Cognitive-Constructivism, Quine, Dogmas of Empiricism, and M\"{u}nchhausen's Trilemma. In: Adriano Polpo, Francisco Louzada, Laura L. R. Rifo, Julio M. Stern, and Marcelo Lauretto (eds.). Interdisciplinary Bayesian Statistics: EBEB 2014, 
\textit{Springer Proceedings in Mathematics \& Statistics}, 118, 55-68.

\rr Stern, Julio Michael  (2017a). Continuous versions of Haack's Puzzles: Equilibria, Eigen-States and Ontologies. 
\textit{Logic Journal of the IGPL}, 25, 4, 604-631.

\rr Stern, Julio Michael  (2017b). Jacob's Ladder: Logics of Magic, Metaphor and Metaphysics: Narratives of the Unconscious, the Self, and the Assembly. \textit{Sophia}, online, \\ \textit{doi:10.1007/s11841-017-0592-y}  
   
\rr Stern, Julio Michael (2018a). Verstehen (causal/interpretative understanding), Erkl\"{a}ren (law-governed description/prediction), and Empirical Legal Studies. 
\textit{Journal of Institutional and Theoretical Economics}, 174, 105-114. 
%\textit{doi:10.1628/093245617X15120238641866} 

\rr Stern, Julio Michael  (2018b). Karl Pearson on Causes and  Inverse Probabilities: Renouncing the Bride, Inverted Spinozism and Goodness-of-Fit. 
\textit{South American Journal of Logic}, 4, 1, 219-252.

\rr Stern, Julio Michael; Izbicki, Rafael; Esteves, Luis Gustavo; Stern, Rafael Bassi (2017). 
Logically-Consistent Hypothesis Testing and the Hexagon of Oppositions.
\textit{Logic Journal of the IGPL}, 25, 741–757. 

\rr Stern, Julio Michael; Pereira, Carlos Alberto de Bragan\c{c}a (2014).  
Bayesian Epistemic Values: Focus on Surprise, Measure Probability! 
\textit{Logic Journal of the IGPL}, 22, 236-254.

%\rr St\"{o}ltzner, Michael (2003). The Principle of Least Action as the Logical Empiricist's Shibboleth. \textit{Studies in History and Philosophy of Modern Physics}, 34, 285-318.  

%\rr St\"{o}ltzner. Michael (1999) To What Extent does Formal Teleology Still Make Sense?. In: Feh\'{e}r M., Kiss O., Ropolyi L. (eds) Hermeneutics and Science. Boston Studies in the Philosophy of Science, v.206. Springer, Dordrecht. 
  
\rr Szab\'{o}, \'{A}rp\'{a}d (1978). \textit{The Beginnings of Greek Mathematics}. Budapest: Akad\'{e}miai Kiad\'{o}. 

\rr Terekhovich, Vladislav (2012). 
\textit{Metaphysics of the Principle of Least Action}. 
\textit{arXiv:1511.03429}  

\rr Texereau, Jean (1957). \textit{How to Make a Telescope}. New York: Interscience. 

\rr Tobin, William (1993). Toothed Wheels and Rotating Mirrors. 
\textit{Vistas in Astronomy}, 36, 253-294. 
 
\rr Tomasello, Michael;  Carpenter, Malinda;  Call, Josep;  Behne, Tanya; Moll, Henrike (2005). Understanding and sharing intentions: The origins of cultural cognition. \textit{Behavioral and Brain Sciences}, 28, 5, 675-735.     

\rr Thompson, Allyn (1947). \textit{Making your own telescope}. Cambridge, MA: Sky Publishing.  

%\rr Trubody (2016). Richard Feynman's Philosophy of Science. \textit{Philosophy Now}, 114, on-line. 
%\textit{https://philosophynow.org/issues/114/Richard\_Feynmans\_Philosophy\_of\_Science} 
      
%\rr Vannoni, Maurizio; Molesini, Giuseppe; Sordini, Andrea; Straulino, Samuele (2011). A Mechanical Analogue of the Refracting Telescope. \textit{Physics Teacher}, 49 4, 236-237.  

\rr Varela, Francisco J. (1979). \textit{Principles of Biological Autonomy}. Amsterdam: North-Holland. 
  
\rr Vasconcellos, D\'{e}cio Fernandes de (1960). \textit{Polyopticon} PO-M1. S\~{a}o Paulo: D.F.Vasconcelos.    
 % Lens kit (focal length): 1 Convergent 19.5mm, 
 % 1 Double achromatic 36.7mm, 2 Convergent 41.8mm, 
 % 2 Divergent 49.5mm, 1 Achromatic objective 305.7mm, 
 % 2 Achromatic objective 155.2mm.      

\rr  Venables, Keith (2017). Roots of Amateur Astronomy. \textit{Sky and Telescope}, 2017, 5, 22-27. 

%\rr  Walter, Gilbert G. (1994). \textit{Wavelets and Other Orthogonal Systems}. Boca Raton: CRC. 

\rr Wang, Qing; Ronneberger, Olaf; Burkhardt, Hans (2008). \textit{Fourier Analysis in Polar and Spherical Coordinates}. Tech. report 1/08, Institut f\"{u} Informatik, Albert Ludwigs Universit\"{a}t Freiburg. 

\rr Welch, Claude A.; Aron, Daniel I.; Cochran, Harold; Erk, Frank C.; Fishlerer, Jack; Mayer, William V.; Pius, Sister M.; Shaver, John; Smith, Frank W. (1968). \textit{Biological Science: Molecules to Man}, BSCS - Biological Science Curriculum Study, revised ed. NY: Houghton Mifflin.

%\rr Wilber, Heather; Townsend, Alex; Wright, Grady B. (2016). Computing with Functions in Spherical and Polar Geometries I. The Sphere. \textit{SIAM Journal on Scientific Computing}, 38, 4, C403-C425. 
  
%\rr Wilber, Heather; Townsend, Alex; Wright, Grady B. (2017). Computing with Functions in Spherical and Polar Geometries II. The Disk \textit{SIAM Journal on Scientific Computing}, 39, 3, C238-C262. 
     
%\rr Wilber, Heather (2017). Eigenfunctions of the Laplacian on the Disk. \\ \textit{www.chebfun.org/examples/disk/Eigenfunctions.html}  	
	   
% \rr Williams, Michael R., Erwin Tomash (2003). The Sector: Its History, Scales, and Uses. \textit{IEEE Annals of the History of Computing}, 25, 1, 34-47.

\rr Wimp, Jet (1984). \textit{Computation with Recurrence Relations}. London: Pitman.  

\rr  Woodfleld, Andrew (1976). \textit{Teleology}.  Cambridge University Press

\rr Yourgrau, Wolfgang; Mandelstam, Stanley (1960). \textit{Variational Principles in Dynamics and Quantum Theory}. NY: Pitman. 
           
\rr Zeleny, Milan (1980). \textit{Autopoiesis, Dissipative Structures, and Spontaneous Social Orders}. Washington:
American Association for the Advancement of Science. 

\rr Zempl\'{e}n , G\'{a}bor A. (2005).  
\textit{The History of Vision, Colour, and Light Theories}. Bern: Gerd Grasshoff. 
  
%\rr Zernike, Frits (1934). Beugungstheorie des Schneidenverfahrens und Seiner Verbesserten Form, der Phasenkontrastmethode. \textit{Physica}, 1, 8, 689-704. 

%\rr Zik, Yaakov; Giora Hon (2017). History of science and science combined: solving a historical problem in optics: The case of Galileo and his telescope. \textit{Archive for History of Exact Sciences}, 71, 4, 337-344.
   
\rr Zou, Ji-Ping; Wattellier, Benoit (2012).  Adaptive optics for high-peak-power lasers–an optical adaptive closed-loop used for high-energy short-pulse laser facilities: laser wave-front correction and focal-spot shaping. Ch.5 in Tyson, Robert K. (2012). \textit{Topics in Adaptive Optics}. Intech, Rijeka, Croatia.

 \end{small} 
 %\end{footnotesize} 
 %\end{scriptsize} 

 %\pagebreak 

\section*{Appendix A: Kepler's Paraxial Optics, \\ 
	Presented in Modern Matrix Notation}  

When analyzing the behavior of a given system, it is often a good idea to try an input-output approach combined with linear approximations. An input-output approach seeks a descriptive functional relation between relevant input and output variables for the system in study.  
In a linear approximation, each output variable is described as a linear combination of the input variables or, equivalently, the input and output vector variables, 
$v$ and $w$,  are related by a linear (matrix) operator, 
$w= Av$. 
In this appendix we follow Kleinfeld and Tsai (2004), see also Pedrotti et al. (2017, Ch.18), and Gerrard and Burch (1975, Ch.1 and 2).

In the Transfer (or propagation) Matrix approach to optics, the optical system is described by the trajectory of light rays in the system, see Figure 18. 
The coordinate $x$ locates the ray as it passes through the system, and vector variables  
$\left[ \begin{array}{cc} y & \alpha \end{array} \right]^{t}$ 
describe the ray's height and the angle (in radians) 
relative to axis $x$.

The action of an optical elements is modeled by a linear 
operator, represented by a matrix $A$, that is,       
\[ 
\left[ \begin{array}{c} y' \\ \alpha' \end{array} \right] = 
A \left[ \begin{array}{cc} 
y \\ \alpha \end{array} \right] = 
\left[ \begin{array}{cc} 
A_{1,1} & A_{1,2} \\ 
A_{2,1} & A_{2,2} \end{array} \right]  
\left[ \begin{array}{cc} 
y \\ \alpha \end{array} \right] =  
\left[ \begin{array}{cc} 
A_{1,1} y +A_{1,2} \alpha \\ 
A_{2,1} y +A_{2,1} \alpha \end{array} \right] 	
\ . 
\] 

Taking the standard basis 
$\left[ \begin{array}{cc} 1 & 0 \end{array} \right]^t$  and 
$\left[ \begin{array}{cc} 0 & 1 \end{array} \right]^t$
as input vectors, we notice that the first column of $A$ 
describes the action of the optical element over 
a light ray parallel to $x$ at unitary height, 
while the second column of $A$ 
describes the action of the optical element  
at zero height and unitary inclination. 

Moreover, in the transfer matrix approach to optics, $C$, the transfer matrix of a system composed by two  optical elements, $A$ and $B$ (aligned left to right), 
is obtained by multiplying (right to left)  the matrices corresponding to its constituent elements.    
\[ 
C = BA = 
\left[ \begin{array}{cc} 
B_{1,1}A_{1,1}  +B_{1,2}A_{2,1} & 
B_{1,1}A_{1,2}  +B_{1,2}A_{2,2} \\  
B_{2,1}A_{1,1}  +B_{2,2}A_{2,1} & 
B_{2,1}A_{1,2}  +B_{2,2}A_{2,2} 
\end{array} \right]
\] 

Linear ray optics offers good approximations 
to real optical systems as long as the following 
hypotheses hold:       
(1) Small angles, that is, $\theta << 1$, so that 
$\sin \theta \approx \tan \theta \approx \theta$;  
(2) Small heights (paraxial rays), that is, $y << r$, 
where $r$ is the radius of spherical elements to be 
specified in the sequel; 
(3) Thin elements, that is, the change in height of a 
light ray while traveling inside an optical element 
is negligible. 
Assuming the aforementioned conditions to hold, 
we derive the transfer matrices for a few optical elements 
and simple optical systems.

\begin{figure}[bt] 
	%\mbox{} \vspace{0mm} \mbox{}     
	
	\centerline{
		\includegraphics[height=1.2in]{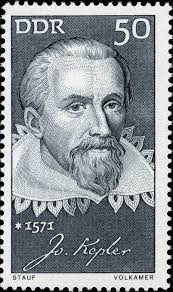} 
		\mbox{} \ 
		\includegraphics[height=1.2in]{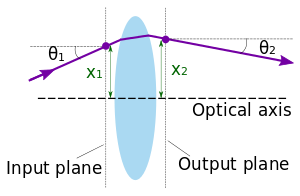} 
		\mbox{} \ \ 
		\includegraphics[height=1.2in]{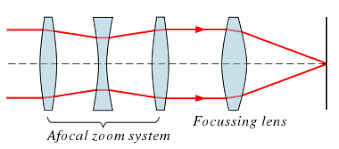} } %width=3.0in
	
	\caption{Keplerian or paraxial optics}  
	
\end{figure}

$D(d)$ -- Distance $d$ simple translation: 
A light ray traveling through an homogeneous medium follows a straight line. Traveling a distance $d$ at angle $\alpha$  
changes the ray's height by $d\tan \alpha \approx d \alpha$. 
\\  \indent 
$F(n,n')$ -- Flat (orthogonal to $x$) interface from medium 
$n$ to $n'$. Crossing a flat interface causes no change in 
the height of a ray, while its angle changes according to 
the linear version of Snell-Descartes law, that is, 
$\sin \theta' = (n/n') \sin \theta \Rightarrow$ 
$\theta' \approx (n/n') \theta$. 
\[ 
D(d) = \left[ \begin{array}{cc} 
1 & d \\ 0 & 1 \end{array} \right]
\ ; \ \ 
F(n,n') = 
\left[ \begin{array}{cc} 
1 & 0 \\ 0 & n/n'   
\end{array} \right] \ . 
\]

$S(n,n',r)$ -- Spherical interface or radius $r$ (where $r$ is positive/ negative  for a convex/ concave interface). 
First, notice that a ray of zero height an angle $\alpha$ 
behaves as passing through a flat interface, giving us 
the second column of matrix $S$. 
Meanwhile, at the point a ray with unit height and  
parallel ($\alpha=0$) to the axis meets the interface, 
the angle between incoming ray and the spherical radius is 
$\theta =\arcsin 1/r \approx 1/r$. 
Using the linear version of Snell's law, we get 
$\theta'= (n/n')\theta$. 
Finally, using the identity 
$\alpha'= \theta' -\theta$, we get 
\[ 
S(n,n',r) = \left[ \begin{array}{cc} 
1 & 0 \\ (n -n')/ r\,n' & n/n'   
\end{array} \right] 				
\ .  
\]

$M(r)$ -- Mirror of radius $r$ (where $r$ is 
positive/ negative  for a concave/ convex spherical mirror). 	
Arguments similar to the ones used in previous case,  
give us the transfer matrix for the spherical mirror. 
\[ 
M(r) = \left[ \begin{array}{cc} 
1 & 0 \\ 2/r & 1   
\end{array} \right] 				
\ .  
\]

We can now obtain the transfer matrices of some 
simple optical systems by multiplying (right to left) 
the transfer matrices of their constituent elements 
(left to right). 

$L(n,n',r_1,r_2)$ -- Lens made by a thin sequence 
of convex and concave interfaces ($r_1>0$ and $r_2<0$), 
with exterior medium $n$ and internal medium $n'$. 
It is easier to write this matrix as $L(f)$, 
using as an auxiliary variable the lens' focal distance,  
$f$, defined as 
$(1/f)= (1/r_1 -1/r_2)(n'-n)/n$,       
\[ 
L(n,n',r_1,r_2) = S(n',n,r_2) S(n,n',r_1) = 
\left[ \begin{array}{cc} 
1 & 0 \\ -1/f & 1 \end{array} \right]
%\ , \ \ \frac{1}{f}= \frac{n' -n}{n} 
%    \left( \frac{1}{r_1} -\frac{1}{r_2} \right) 
= L(f) \ .  
\]

$T(f_1,f_2,d)$ -- Telescope using two thin lenses with focal 
distances $f_1$ and $f_2$ separated by a distance $d$.  
\[ 
T(f_1,f_2,d) = L(f_2) L(f_1) = 
\left[ \begin{array}{cc} 
1 -d/f_1 & d \\ 
(d -f_1 -f_2)/ (f_1\,f_2) & 1 -d/f_2 \end{array} \right] 		
\ . 
\] 

$C(f_1,f_2)$ -- Collimated telescope, namely, 
$T(f_1,f_2,d)$ adjusting $d=f_1 +f_2$.     
The matrix element $A_{1,1}= -f_2/f_1$ gives the collimated 
telescope's magnification coefficient for the image of a 
distant object; the negative sign indicating an inverted image. 
\[    
C(f_1,f_2) = \left[ \begin{array}{cc} 
-f_2/f_1 & f_1 +f_2 \\ 
0 & -f_1/f_2 \end{array} \right] 		
\ .  
\]  

 %\pagebreak 

\section*{Appendix B: Principle of Least Time} 

Consider a beach with shore line represented by $x=a$, in the standard 
Cartesian plane. 
A lifeguard, at position $(x,y)=(0,0)$, spots a person drowning at
position $(x,y)=(a+b,d)$. 
While on the athletic track the lifeguard car can run at top speed 
$c$, on the sand it can run at speed $c/\nu_1$. 
Once in the water, the lifeguard can only swim at speed 
$c/\nu_2$, $1< \nu_1< \nu_2$. 
Letting $(x,y)=(a,y(a))$ be the point where he enters the water, what
is the optimal value $y(a)=z$ if he wants to reach position $(a+b,d)$ as
fast as possible?

Since the shortest path in an homogenous medium is a straight line, the
optimal trajectory is a broken line, from $(0,0)$ to $(a,z)$, and then
from $(a,z)$ to $(a+b,d)$. The total travel time is $J(z)/c$, where
\[
J(z)= \nu_1 \sqrt{a^2+z^2} +\nu_2 \sqrt{b^2+(d-z)^2} \ .
\]
Since we want $J(z)$ at a minimum, we set
\[
\frac{dJ}{dz}=
\nu_1 \frac{-2z}{2\sqrt{a^2+z^2}}
+\nu_2 \frac{-2(d-z)}{2\sqrt{b^2+(d-z)^2}} =0 \ ,
\]
so that, we should have
\[
\nu_1 \sin(\theta_1) = \nu_2 \sin(\theta_2) \ .
\]

Professional lifeguards claim that this simple model can be improved by
dividing the sand in a dry band, $V_1$, and a wet band, $V_2$, and the
water in a shallow band, $V_3$, and a deep band, $V_4$, with respective
different media `resistance' indices, 
$\nu_1 , \nu_2 , \nu_3 , \nu_4$, satisfying 
$\nu_4> \nu_3> \nu_1 >\nu_2 >1$. 
Although the solution for the improved model can be similarly obtained, a general formalism to solve `variational' problems of this kind exists which is known as the Euler-Lagrange equation. For intuitive
introductions see Krasnov et al. (1973), Leech (1963) and Marion (1970).

%\clearpage  

%\mbox{} 

\end{document}